\documentclass[a4paper,fleqn,usenatbib]{mnras}
\usepackage{amsmath}
\usepackage{amsfonts}
\usepackage{amsbsy}
\usepackage{amssymb}
\usepackage{amsbsy}
\usepackage{makecell}
\usepackage[T1]{fontenc}
\usepackage[utf8]{inputenc}
\usepackage{ae,aecompl}
\usepackage{subfigure}
\usepackage{verbatim}
\usepackage{graphicx}
\usepackage{color}
\usepackage{multirow}
\usepackage{url}
\usepackage{lastpage}

\newcommand{\pD}[2]{\frac{\partial #2}{\partial #1}}

\newcommand\bb[1]{\mbox{\boldmath{$#1$}}}
\newcommand\grad{\bb{\nabla}}
\newcommand\bcdot{\,\bb{\cdot}\,}
\newcommand\btimes{\,\bb{\times}\,}
\newcommand{\msb}[1]{\bb{\mathsf{#1}}}

\newcommand{\treta}{{\rm tr}(\bb{\eta})}
\newcommand{\deteta}{{\rm det}(\bb{\eta})}
\newcommand{\imag}{{\rm i}}
\newcommand{\rmd}{{\rm d}}

\newcommand{\ek}{\hat{\bb{k}}}
\newcommand{\vasq}{v^2_{\rm A}}
\newcommand{\kva}{(\bb{k}\bcdot\bb{v}_{\rm A})}
\newcommand{\khatva}{(\ek\bcdot\bb{v}_{\rm A})}
\newcommand{\ex}{\ensuremath{\bb{e}_{x}}}
\newcommand{\ey}{\ensuremath{\bb{e}_{y}}}
\newcommand{\ez}{\ensuremath{\bb{e}_{z}}}
\newcommand{\ephi}{\ensuremath{\bb{e}_{\phi}}}

\defcitealias{lp18}{LP18}
\defcitealias{cl21}{CL21}

\title[The VSI in non-ideal MHD]
 {The vertical shear instability in poorly ionised, magnetized protoplanetary discs}
\author[H.~N.~Latter \& M.~W.~Kunz]{
Henrik N.~Latter$^{1}$\thanks{Contact e-mail: \href{mailto:hl278@cam.ac.uk}{hl278@cam.ac.uk}} and  Matthew W.~Kunz$^{2,3}$ 
\\
$^{1}$Department of Applied Mathematics and Theoretical Physics,
University of Cambridge,
Wilberforce Road, Cambridge CB3 0WA, UK. \\
$^{2}$Department of Astrophysical Sciences, Princeton University, Peyton Hall, Princeton, NJ 08544, USA\\
$^{3}$Princeton Plasma Physics Laboratory, PO Box 451, Princeton, NJ 08543, USA
 }

\date{Accepted 2022 January 9. Received 2021 December 22; in original form 2021 September 21}

\pubyear{2021}

\begin{document}
\label{firstpage}
\pagerange{\pageref{firstpage}--\pageref{lastpage}}
\maketitle

\begin{abstract}
Protoplanetary discs should exhibit a weak vertical variation in their rotation profiles. Typically this `vertical shear' issues from a baroclinic effect driven by the central star's radiation field, but it might also arise during the launching of a magnetocentrifugal wind.
As a consequence, protoplanetary discs are subject to a hydrodynamical instability,
the `vertical shear instability' (VSI), 
whose breakdown into turbulence could transport
a moderate amount of angular momentum and facilitate, or interfere with, the
process of planet formation.
Magnetic fields may suppress the VSI, however, either directly via magnetic
tension or indirectly through magnetorotational turbulence. 
On the other hand, protoplanetary discs exhibit notoriously low ionisation fractions, and non-ideal effects, if sufficiently
dominant, may come to the VSI's rescue. 
In this paper we develop a local linear theory that explores how non-ideal MHD influences the VSI, while also launching additional diffusive shear instabilities. We derive a set of analytical criteria that establish when the VSI prevails, and then show how it can be applied to a realistic global model of a protoplanetary disc. Our calculations suggest that within ${\sim}10~{\rm au}$ the VSI should have little trouble emerging in the main body of the disk, but beyond that, and in the upper regions of the disc, its onset depends sensitively on the size of the preponderant dust grains.
 
\end{abstract}


\section{Introduction}

Owing to their low ionisation fractions, it is now well established that the magnetohydrodynamics (MHD) of protoplanetary (PP) discs is remarkably complicated, and that the magnetorotational instability (MRI) is likely absent or very sluggish at most disc radii \citep[e.g.,][]{turner14,lesur20}. This predicament has renewed interest in various hydrodynamic instabilities \citep[subcritical baroclinic instability, convective overstability, vertical shear instability, etc.;][]{fl19,lu19}, especially in the later Type II evolutionary stage when angular-momentum transport by spiral density waves has dropped off \citep[e.g.,][]{balbus03,lesur15,kratter16}. Though these instabilities appear collectively to drive only negligible levels of accretion, they likely have a much larger impact on dust settling, radial drift, and coagulation.

The disc's poor ionisation does not mean that magnetic fields can be neglected out of hand, however: recent work has shown its critical role in generating outflows and large-scale fields \citep{bs13,lkf14,bai14,gressel15,simon15},
ringlike coherent structures \citep{kl13,bethune16,bethune17,krapp18,suriano18,suriano19}, and weak diffusive instabilities (e.g., the ambipolar-diffusion shear instability, or ADSI; \citealt{kunz08}). One area that remains unexplored is the interplay between non-ideal MHD and the hydrodynamical instabilities
mentioned above. Are these relationships inimical or enabling? 

Of the set of instabilities discussed, the vertical shear instability (VSI) appears to be one of the most robust and prevalent.
As its name suggests, it draws energy from any vertical shear in the disc, though
in essence it is a centrifugal instability, akin to the Goldreich--Schubert--Fricke
instability of stellar interiors \citep{gs67,fricke68}. A gentle vertical variation in the rotation rate arises generically in PP discs due to irradiation by the protostar
 \citep{ks86,ub98,nelson13}. Meanwhile, steady magnetic outflows produce significant vertical shear acting within the disc, as demonstrated
by \citet{ogilvie97} and \citet{ol01} in ideal MHD, and by \citet{salmeron11}, \citet{bethune17} and  others in non-ideal MHD. Both manifestations of vertical shear, under certain circumstances, are unstable to the VSI. 

The ensuing VSI turbulence has been tracked with global simulations and exhibits a number of interesting features, such the emergence of large-scale inertial wavetrains, zonal flows, and vortices \citep[e.g.,][]{nelson13,sk14,richard16,stoll17}. The radial angular-momentum transport measured is only modest, as the VSI mainly mixes angular momentum vertically, but the large-scales structures it generates might be crucial for the concentration and coagulation of solid particles \citep{sk16,lp18,flock20}.

Almost all of the work cited above treats the PP disc as purely hydrodynamical, but \citet[hereafter \citetalias{lp18}]{lp18} showed (in a simple Boussinesq model) 
that magnetic tension can easily stabilise the VSI. 
A rough instability criterion 
in ideal MHD is that $\beta \gtrsim [R (\partial\ln\Omega/\partial z)]^{-2}\sim (R/H)^2 \gg 1$, where $\beta$ is the plasma beta parameter, $\Omega$ is the orbital frequency, $R$ is the cylindrical radius, and $H$ is the disc scale height. In other words, only rather weak fields
permit the development of the VSI, especially at the higher disc altitudes it favours.
Indeed, the weak turbulent line-broadening observed
in these upper layers \citep{flaherty15,flaherty17,flaherty18} appears consistent with the magnetic suppression of the VSI far off the midplane. Similarly, the VSI might also struggle in the better ionised gas at larger disc radii.
Of course,
ideal MHD is a poor model for PP discs and so these results need to be generalised to their appropriate non-ideal regimes. This constitutes the main motivation for our paper.

We adopt a local incompressible model for a PP disc, as it supplies relatively convenient and consistent results. We examine the VSI in the presence of a vertical and azimuthal magnetic field adding Ohmic diffusion, ambipolar diffusion, and the Hall effect, one at a time. Our focus is on parameters for which the MRI is completely stable or is expected to saturate in a laminar magnetic state; if MRI turbulence is present we assume that it would likely overwhelm the VSI. We find that if any one of the non-ideal effects is sufficiently strong 
then the stabilising influence of magnetic tension is subverted and hydrodynamic results recovered. In addition, when azimuthal magnetic fields are present, the ADSI can merge with the VSI into a hybrid instability. We derive and collate a set of easy-to-apply criteria that can determine whether the VSI prevails; these constitute the main results of the paper.
Finally, we construct realistic ionisation models for a minimum mass solar nebula (MMSN) at different locations and assess the preponderance of the VSI/ADSI. It turns
out that instability is expected at radii less than ${\sim}10~{\rm au}$ and at vertical locations ${<}H$, but that the VSI struggles at larger radii. However, if there are sufficiently small dust grains (${\lesssim}0.1~\mu{\rm m}$) then the VSI region could extend significantly farther. 

The structure of the paper is as follows. In Section \ref{sec:prelim} we provide a brief summary of the main instabilities that appear in this paper -- the VSI, the ADSI, the Hall shear instability (HSI), and a Hall-modified version of the MRI. Section \ref{sec:eqns} then introduces the local incompressible approximation that we adopt, its governing equations, and the general dispersion relation that linear disturbances in the disc must obey. We analyse this dispersion relation in Section \ref{sec:analysis}, investigating each non-ideal case separately, obtaining stability criteria, asymptotic growth rates, and full numerical growth rates. The criteria for VSI emergence are summarised in Section \ref{CriteriaSummary}. We next apply these results in Section \ref{sec:ppdisc} to various locations in a representative PP disc model and assess the prevalence and strength of the VSI and the diffusive shear instabilities. Our conclusions are drawn in Section \ref{sec:conclusion}.

\section{Preliminaries}\label{sec:prelim}

In this section we describe the basic physics of the VSI and the non-ideal-MHD shear instabilities, reflecting the expositions in \citet{bl15}, \citet{kunz08}, and \citet{bt01}. Though the classical MRI is an important ingredient in our paper's calculations, we do not spend any time explaining it here; it is relatively well understood in the field.
Our aim is to be as non-mathematical as possible, and instead to elucidate in an intuitive way the underlying instability mechanisms in each case. 
Readers familiar with these instabilities may wish to skip this section and proceed directly to Section \ref{sec:eqns}.

\subsection{The vertical shear instability (VSI)}

\subsubsection{Equilibrium vertical shear}

Baroclinic disc equilibria generate vertical shear. Consider the ‘thermal wind equation’, the azimuthal component of the vorticity equation
for an axisymmetric, hydrodynamic disc in equilibrium. It can be written as
\begin{equation}\label{eqn:thwind}
\partial_z (R\Omega^2) = -\ephi \bcdot (\grad\rho\btimes\grad P)/\rho^2 =\partial_R T \partial_z S - \partial_z T \partial_R S .
\end{equation}
Here we have adopted cylindrical polar coordinates centred on the central object, $(R, \phi, z)$; $\Omega$ is the orbital frequency; $\rho$, $P$, and $S$ are, respectively, the density, pressure, and entropy of the disc's equilibrium, and the last equality assumes a perfect gas equation of state. 
Equation \eqref{eqn:thwind} states that if the disc equilibrium is baroclinic, i.e. pressure and density do not share the same spatial profile, then the right hand side forces $\Omega$ away from cylindrical rotation. Importantly, this holds true in PP discs on intermediate to large radii, as they are thought to be locally isothermal  \citep[e.g.,][]{dalessio}: $T$ varies with $R$ but not $z$. As a result, the first term in the last equality is non-zero because entropy must increase with height. An estimate for the size of this vertical shear is then $\partial_z(R\ln \Omega)\sim H/R \ll 1$.    

Alternatively, if poloidal magnetic fields feature in the equilibrium balance, the rotation profile exhibits vertical shear via Ferraro's isorotation law (see \citealt{ogilvie97} for various examples). Indeed, shearing velocity 
configurations are essential if the disc is to support a steady magnetocentrifugal 
wind \citep{ol01}. While most prominent in ideal MHD, such vertical shear also appears to varying extents when non-ideal MHD effects are included \citep[e.g.,][]{salmeron11,gressel15,bethune17}.

\subsubsection{Underlying centrifugal mechanism of instability}\label{sec:VSIphysics}

Consider a fluid ring at $(R_0,z_0)$ with angular momentum $\ell_0=R_0^2\Omega_0$, where $\Omega_0=\Omega(R_0,z_0)$. Suppose the ring is slightly displaced by $\delta\bb{s}=(\delta R, \delta z)$ but retains its angular momentum. The centrifugal acceleration it experiences at its new location is $\ell_0^2/(R_0+\delta R)^3$, while the inward acceleration that resists it is $R_0\Omega_0^2+\delta(R\Omega^2)$. If we expand both expressions in small $\delta R$ and $\delta z$, then the force imbalance is proportional to $-\delta\bb{s}\bcdot(\grad \ell^2)_0$ to leading order. If this imbalance is positive, then the initial radial perturbation is exacerbated and instability proceeds. This is always possible for sufficiently small $\delta R/\delta z$ (so that the stabilising effect of the radial $\ell$ gradient is minimised) and if $\delta z (\partial_z\ell^2) <0$. But as long as there is some vertical variation in the angular momentum (i.e.~$\partial_z\Omega \neq 0$) we can choose the sign of $\delta z$ appropriately to achieve instability. Note that it is possible to construct a separate Rayleigh-type argument, involving the swapping of two fluid rings, that reinforces this conclusion and shows explicitly that energy is liberated by the swap.


In summary, though accretion discs are stable
according to the classical Rayleigh's criterion, any vertical
shear at all permits its circumvention and hence the onset of
instability. For weak shear, as is expected (see previous Section), the instability prefers a poloidal wavevector oriented predominantly in the radial direction, as the displacements (discussed above) are vertically elongated with $\delta R/\delta z \ll 1$.

\subsubsection{Double-diffusive aspects}

For the same reason that vertical shear appears in PP discs, so must stable stratification. Thus our physical picture of instability needs to account for buoyancy forces. In particular, buoyancy opposes displacements of the type described above and thus inhibits any adiabatic (dynamical) instability (cf.\ the
Solberg--H\o iland criterion). But if the buoyancy forces can be eliminated, such as by sufficiently fast
 thermal diffusion, then instability is restored.
For this to work, displacements $|\delta \bb{s}|$ must be much
 shorter than the thermal diffusion scale. 
 The resulting instability is hence double-diffusive in character, possessing
 wavelengths lying in a range bounded from below by the
 viscous length and above by the thermal diffusion length.
 
This characterisation is accurate, however, only where the photon mean free path is less than the wavelengths of the unstable modes -- and thus the diffusion approximation acceptable. This `optically thick' regime encompasses inner disc radii within ${\sim}10~{\rm au}$, though can vary appreciably depending on the global disc profiles adopted (\citealt{ly15}; \citetalias{lp18}). In the optically thin gas outside this region it is typical to model the disc cooling using a single, possibly $R$-dependent, cooling time for \emph{all} lengthscales \citep[e.g.,][]{malygin17,pk19}, in which case a sufficiently fast cooling can negate the stable stratification. But it should be stressed that this approximation should not be adopted in the inner optically thick disc, as it prohibits the VSI seeking out the preferred scale upon which it grows at its maximum rate. Models that use an optically thin cooling law
 generally underestimate the prevalence and vigour of the VSI at smaller radii. 

\subsection{The ambipolar-diffusion shear instability (ADSI)}\label{sec:adsi}

In systems with a low degree of ionisation, the dominant neutral species can drift with respect to the charged species (and thus the magnetic field) in a process known as ambipolar diffusion. Under conditions relevant to molecular clouds, protostellar cores, and young PP discs, this drift is determined mainly by a quasi-static balance between the Lorentz force and collisional drag, both acting on the charged species (the latter being mainly due to ion--neutral collisions). By Newton's third law, this drag indirectly communicates the presence of the magnetic field to the neutral species. As a result, the magnetic field can be considered as partially frozen into the dominant neutral species but for a nonlinear diffusion that is proportional to the Lorentz force divided by the neutral--ion collision frequency.

From the standpoint of this paper, the most important feature of ambipolar diffusion is not its nonlinear nature, but rather that it acts {\em anisotropically} with respect to the local magnetic-field direction. That is, only the component of the current density perpendicular to the magnetic field is damped, since it is this component that is responsible for forcing the ions (into which the field is assumed to be frozen) through the neutrals. In practice, this means that perturbations to the magnetic field tend to align themselves to be perpendicular to the background magnetic field, especially so for wavevectors with large field-perpendicular components. 

In the presence of velocity shear, and for suitably oriented wavevectors and magnetic fields, this can lead to a feedback loop whereby the shear stretches the transverse field components into the streamwise direction, with ambipolar diffusion effectively projecting that streamwise component back into the transverse direction to be sheared further (a graphical depiction of this process can be found in fig.~7 of \citealt{kunz08}). This is the essence of the ADSI: preferentially perpendicular current damping by ambipolar diffusion conspires with shear to produce exponential growth of magnetic perturbations (albeit at a small fraction of the shear frequency).

When the shear is instead supplied by a disc's differential rotation, the accompanying Coriolis force has a stabilizing influence on the ADSI by inducing epicylic motion in the neutrals -- an influence that can be partly mitigated by orienting the poloidal wavevector predominantly in the radial direction. Given the VSI's strong preference for similarly oriented wavevectors, it is natural to expect that the two instabilities interact. Finally, for the vertically oriented wavevectors that are more customarily considered \citep[e.g.,][]{bb94}, ambipolar diffusion simply damps linear magnetorotational perturbations in a way akin to Ohmic dissipation.

\subsection{The Hall shear instability (HSI) and Hall MRI}\label{sec:hsi}

A similar shear instability afflicts systems in which the heavier ion species drifts appreciably with respect to the flux-frozen electrons, a drift generally referred to as the Hall effect. For Alfv\'enic disturbances in an ion--electron plasma, the Hall effect becomes important on lengthscales comparable to the ion skin depth $d_i$, below which such disturbances become affected by the gyromotion of the ions. In a poorly ionised plasma, this critical lengthscale is augmented by the inertia of the neutrals to which the ions are collisionally coupled, boosting what is customarily considered a microscopic scale ($d_i$) by a large factor of $(\rho/\rho_i)^{1/2}$, where $\rho$ denotes the mass density of the neutrals and $\rho_i$ that of the ions \citep[see, e.g.,][]{pw08}. Considering the low degrees of ionisation in protostellar cores and PP discs, the Hall effect may then become important on macroscopic (`fluid') lengthscales.

For our purposes, it is useful to sort the impact of the Hall effect on disc stability into two categories: the HSI, and a Hall-modified version of the MRI. First, the HSI. As with the ADSI, the HSI depends on a feedback loop between velocity shear, which stretches transverse magnetic fields in the streamwise direction, and anisotropic diffusion, which re-orients those sheared fields back into the transverse direction. But instead of the latter arising from a dissipative projection, the Hall effect brings about a conservative rotation: at scales below $\ell_{\rm H}\equiv d_i (\rho/\rho_i)^{1/2}$, Hall electromotive forces induce a handedness to the field perturbations that, for certain wavevector orientations, pivots those fields back into the transverse direction without dissipation to be sheared further still (a graphical depiction of this process can be found in fig.~8 of \citealt{kunz08}). The resulting instability holds whether the shear is planar or rotational, though in the latter case the accompanying epicyclic motion enjoys a competitive or cooperative interplay with the Hall-induced circular polarization that effectively lowers or boosts the Coriolis force. The latter affects how responsive the disc is to magnetorotational perturbations \citep{wardle99,bt01}, to which we now turn.

Previous work (omitting vertical shear) has shown that when the background magnetic field and the rotation axis are co-aligned the Hall effect tends to suppress the MRI \citep{wardle99}. In this case, by inducing circular polarization, Hall electromotive forces effectively slow the dynamical epicyclic response and increase the restoring radial magnetic tension force \citep{bt01}. What takes the MRI's place in the highly diffusive limit is the HSI \citep{kunz08,pw12}, which in a differentially rotating disc amounts to a favorably sheared whistler wave. In this case, it is predominantly the right-handedness of the whistler wave, rather than the radially directed magnetic tension force, that generates radial (transverse) magnetic fields at the expense of azimuthal (streamwise) magnetic fields, and thereby completes the feedback loop with the Keplerian shear.

When the magnetic field and rotation axis are counter-aligned, the field-line drift through the fluid caused by the Hall effect undermines the tension force, allowing the destabilizing dynamical tide to operate in what \citet{pw12} refer to as the `diffusive MRI' (DMRI). This is essentially a low-frequency ion-cyclotron oscillation destabilized by in-phase epicyclic motion in the bulk-neutral fluid \citep{ws12,simon15}. In the highly diffusive limit, the usually destabilizing azimuthal torque on the fluid elements becomes stabilizing, as the circular polarization induced by the Hall effect reinforces the stabilizing epicyclic motion to overcome the transfer of angular momentum by magnetic tension \citep{bt01}.

In Section \ref{sec:Hall} we show, by varying the ratio $\ell_{\rm H}/H$ and the orientation of the background magnetic field, that the presence of vertical shear causes the DMRI to merge with the VSI and the HSI to operate off the vertical shear.

\section{Mathematical model and governing equations}\label{sec:eqns}

Having covered the background physics animating the problem, in this section we outline the mathematical model with which we describe it. In summary, our model assumptions and equations are presented in Sections \ref{model} and \ref{eqns}, the background equilibrium disc state is given in Section \ref{sec:equil}, the formulation of the linear theory in Sections \ref{sec:lineareqns} and \ref{sec:disprel}, and important dimensionless free parameters in Section \ref{sec:parameters}.

\subsection{The incompressible shearing sheet}
\label{model}

Given the inevitably complex interactions between the VSI, the MRI, and non-ideal MHD, we adopt the simplest possible physical model containing all the necessary dynamical ingredients, namely  incompressible MHD in the shearing sheet. We assume that perturbations to a background global equilibrium are (i) small-scale, (ii) significantly sub-sonic, and (iii) comprise very small thermodynamic variations. The formal derivation of the governing equations then proceeds via an asymptotic expansion in two small parameters, $\lambda/H$ and $\mathcal{M}$, where $\lambda$ is a typical perturbation lengthscale and $\mathcal{M}$ is the perturbation Mach number. In addition, one adopts the scalings $\delta \rho/\rho_0\sim \delta P/P_0\sim \mathcal{M}^2\ll 1$, where $\delta\rho$ and $\delta P$ are density and pressure perturbations, and $\rho_0$ and $P_0$ are their equilibrium values \citep[see Section 3 in][]{lp17}.

We next assume the fluid to be composed only of ions, electrons, and neutrals, though the formalism would be the same if we were to have included charged grains. While PP discs are poorly ionised, the ions and neutrals are sufficiently well coupled for a single-fluid treatment to be acceptable. In other words, the inertia of the charged species is negligible to that of the neutrals, so that they reach a terminal velocity set by Lorentz and drag forces instantaneously \citep[e.g., see][]{balbus11}.

Two modelling issues are worthy of further discussion. First, by virtue of assuming equally small fractional density and pressure perturbations, thermal physics and buoyancy effects disappear from the leading-order equations. Relaxing this restriction, i.e.~adopting $\delta P/P_0 \ll \delta \rho/\rho_0 \ll 1$, would reinstate this physics and give us the Boussinesq equations in the shearing sheet \citep[see Section 4 in][]{lp17}. We keep to the simpler incompressible model mainly because the analysis is cleaner; however, it should be noted that, within 10~au in less massive disc models, a wide spectrum of wavelengths ${\lesssim}H$ can relax thermally at a rate fast enough to keep the fractional density perturbation small and thus buoyancy effects negligible \citepalias{lp18}.

Second, at leading order in our expansion, the background global equilibrium's thermal gradients drop out, though the orbital shear in radius must be kept. The orbital shear in the vertical direction may also be included consistently, even though it is small, if we make the additional assumption that $\lambda/H \ll H/R\ll 1$. The vertical shear term will be ${\sim}H/R$ smaller than the leading-order terms, but nonetheless much greater than all the omitted terms \citep{lp17}. We are then justified in retaining that subdominant effect.


\subsection{Governing equations}\label{eqns}

Our local model takes the form of a small patch of disc centred at $R = R_0$ and $z = z_0 \neq 0$,
 orbiting with frequency $\Omega_0 = \Omega(R_0, z_0)$, and described using a
co-rotating Cartesian reference frame. In it, the local radial, azimuthal, and vertical directions are represented by the coordinates $x$, $y$, and $z$, respectively, with their origin the centre of the patch. 
Instead of working with the perturbation equations, it is convenient to start with those governing the `total variables', which include the equilibrium fields as they manifest locally under the various assumptions described in the previous subsection. 
The equations of motion and of magnetic induction are then:
\begin{align}
\pD{t}{\bb{v}} &= - \bb{v}\bcdot\grad\bb{v} - \frac{1}{\rho} \grad P +
                 \frac{\bb{J}\btimes\bb{B}}{c\rho} \nonumber\\*
\mbox{} &\quad - 2\bb{\Omega}_0 \btimes \bb{v} + 2\Omega_0^2(q_R x +  q_z z)  \ex , \label{eqn:momentum}
\\
\pD{t}{\bb{B}} &= \grad\btimes\left[ \bb{v}\btimes\bb{B} - \frac{4\pi\eta\bb{J}}{c} - \frac{\bb{J}\btimes\bb{B}}{e n_{\rm e}} + \frac{(\bb{J}\btimes\bb{B})\btimes\bb{B}}{c\rho\nu_{\rm ni}} \right] . \label{eqn:induction}
\end{align}
In addition, there are the solenoidality constraints,
\begin{equation}
\grad\bcdot\bb{v} = 0 \qquad {\rm and}\qquad
\grad\bcdot\bb{B} = 0 , \label{eqn:divB}
\end{equation}
and Amp\`{e}re's law,
\begin{equation}\label{eqn:ampere}
\bb{J} = \frac{c}{4\pi} \grad\btimes\bb{B} .
\end{equation}
The notation used here is standard: $\bb{v}$, $P$, $\bb{J}$, $\bb{B}$ denote the fluid velocity, gas pressure, current density, and magnetic field, respectively, while $\rho$ denotes the (assumed constant) mass density. 
We have also introduced the following dimensionless orbital and vertical shear parameters:
\begin{equation}
q_R \equiv -\pD{\ln R}{\ln \Omega}\biggr|_{R_0,z_0} \quad{\rm and}\quad q_z \equiv - \pD{\ln z}{\ln \Omega}\biggr|_{R_0,z_0} .
\end{equation}
The corresponding (dimensional) shear rates are $A_0 \equiv -(1/2)q_R \Omega_0$ characterising the radial shear and $S_0 \equiv -(1/2)q_z \Omega_0$ characterising the vertical shear. With these definitions, the squared epicyclic frequency is
\begin{equation}
    \kappa^2 = 2\Omega^2_0(2-q_R)=4\Omega^2_0 (1+A_0/\Omega_0) .
\end{equation}
A Keplerian disc possesses $q_R=3/2$, and
typically $|q_z| \sim H/R_0 \ll 1$. (Note that $q_z$ corresponds to the parameter $q$ in \citetalias{lp18}.) The specific angular momentum in the shearing box (i.e., the canonical $y$-momentum) is given by
\begin{equation}
    \ell= (2-q_R) \Omega_0x -  q_z\Omega_0 z = 2\Omega_0 (1+A_0/\Omega_0)x + 2S_0 z.
\end{equation} 
Lastly, the constants $\eta$, $c$, $e$, $n_{\rm e}$, and $\nu_\text{ni}$ that appear in the induction equation \eqref{eqn:induction} denote the Ohmic resistivity, the speed of light,
elementary charge, the number density of electrons, and the
neutral-ion collision frequency, respectively.

\subsection{Equilibrium state}\label{sec:equil}

As explained in \citet{lp17} and \citetalias{lp18}, the background global disc equilibrium manifests in the local incompressible model as a steady homogeneous state of linear shear in both the radial and vertical directions,
\begin{equation}
\bb{v}=\bb{v}_0 = -\Omega_0(q_R x + q_z z) \ey ,
\end{equation}
with constant $P=P_0$ and $\bb{B}=\bb{B}_0$. Of course, globally both $\rho$ and $P_0$ vary spatially, but local subsonic perturbations only encounter this variation indirectly through the $q_z$ term in equation~\eqref{eqn:momentum}.
The components of $\bb{B}_0$ are not arbitrary and must satisfy 
\begin{equation}
\pD{t}{B_{0y}} = -\Omega_0( q_R B_{0x} + q_z B_{0z}) = 0 
\end{equation}
(Ferraro's law of isorotation), which implies
\begin{equation}
\frac{B_{0x}}{B_{0z}} = -\frac{q_z}{q_R} \ll 1 .
\end{equation}
Thus we are free to choose the $y$ and $z$ components of $\bb{B}_0$, but $B_{0x}$ is then determined uniquely by $B_{0z}$ and the shear rates. Finally, it will be convenient in what follows to employ the equilibrium Alfv\'en velocity $\bb{v}_{\rm A} \equiv (B_{0x},\,B_{0y},\,B_{0z})/\sqrt{4\pi\rho}$ and to drop henceforth the subscript `0'.

\subsection{Linearised equations}\label{sec:lineareqns}

We perturb the equilibrium state described in Section \ref{sec:equil} with small perturbations $\delta v_x$, $\delta v_y$, $\delta v_z$, $\delta B_x$, etc. After inserting these into equations \eqref{eqn:momentum} and \eqref{eqn:induction}, linearising in their amplitudes, and assuming  axisymmetry, these perturbations take the form ${\propto}\exp(s t + \imag\bb{k}\bcdot\bb{r})$, where $s$ is a (possibly complex) growth rate and the wavevector
$\bb{k}=(k_x,\,0,\,k_z)$ is real. Note that the adoption of axisymmetry is not especially restrictive as the non-axisymmetric VSI can be shown always to decay \citepalias{lp18}.

The equations for the horizontal components of the velocity and magnetic perturbations may be manipulated into an attractive form, once we introduce the planar velocity perturbations $\delta\bb{v}= (\delta v_x,\,\delta v_y)$ and $\delta\bb{v}_{\rm A} = (\delta B_x,\,\delta B_y)/\sqrt{4\pi\rho}$, and the resistivity tensor $\bb{\eta}$ (defined below):
\begin{align}
    &s\,\delta \bb{v} - 2\Omega\frac{k_z^2}{k^2} \delta v_y\ex +
    \frac{\widetilde{\kappa}^2}{2\Omega} \delta v_x\ey 
   - \imag\kva \delta \bb{v}_{\rm A} = 0, \label{eqn:mom}
    \\
   &s\,\delta \bb{v}_{\rm A}  - \imag\kva \delta \bb{v}
     - \widetilde{S}\,\delta
   v_{{\rm A}x}\ey + k^2 (\bb{\eta}\bcdot\delta\bb{v}_{\rm A}) =0 \label{eqn:ind}.
\end{align}
In writing equations \eqref{eqn:mom} and \eqref{eqn:ind}, we have used the solenoidality conditions \eqref{eqn:divB} to eliminate the vertical components of the perturbed velocity and magnetic field in favour of their horizontal counterparts. We now define the new notation we have introduced.

Following \citet{kunz08}, the (horizontal) resistivity tensor $\bb{\eta}$ possesses components given by
\begin{align}
\eta_{ij} &\equiv \left( \eta + \frac{\vasq}{\nu_{\rm ni}} \right)
\delta_{ij} - \frac{(\ek\btimes\bb{v}_{\rm A})_i
  (\ek\btimes\bb{v}_{\rm A})_j}{\nu_{\rm ni}} \notag \\ 
 & \hskip1cm+ \frac{k_j}{k_z}
\frac{(\ek\btimes\bb{v}_{\rm A})_i (\ek\btimes\bb{v}_{\rm
    A})_z}{\nu_{\rm ni}} + \frac{c\hat{k}_z(\ek\bcdot\bb{B})}{4\pi
  en_{\rm e}}\, \mathsf{H}_{ij} \label{eqn:eta} ,
\end{align}
where the indices only take values of $x$ or $y$ and $\ek = \bb{k}/|k|$ is the unit wavevector. The 2D `Hall matrix' whose elements appear in the final term of equation \eqref{eqn:eta} is 
\begin{equation}
\msb{H} \equiv
\begin{pmatrix}
0 & 1 \\
-k^2/k^2_z & 0
\end{pmatrix} .
\end{equation}
The final (Hall) component of the resistivity tensor is trace-less; when $k=k_z$, it represents a rotation of the magnetic-field perturbation about the $z$ axis.

Equations \eqref{eqn:mom} and \eqref{eqn:ind} feature two modified disc frequencies. The first is an `effective epicyclic frequency' given by
\begin{equation}\label{eqn:axiDR}
    \widetilde{\kappa}^2 \equiv \kappa^2 + 2q_z\Omega^2 \frac{k_x}{k_z}
   =\frac{\bb{k}}{k_z}\bcdot(\grad \ell)_\perp,
\end{equation}
in which $(\grad\ell)_\perp = 2 q_z\Omega^2\ex + \kappa^2 \ez$ is a vector perpendicular to the background angular-momentum gradient. Because disturbances are incompressible, the final equality in equation \eqref{eqn:axiDR} means that $\widetilde{\kappa}^2\propto -\delta\bb{s}\bcdot\grad \ell$, where $\delta\bb{s}$ is the mode's displacement. According to the arguments in Section~\ref{sec:prelim}, a negative $\widetilde{\kappa}^2$ indicates that a displaced fluid blob experiences a force imbalance that exacerbates its initial perturbation.
Thus centrifugal instability occurs when
$\widetilde{\kappa}^2<0$,
that is, when the effective epicyclic frequency is imaginary.

The second modified disc frequency is an `effective shear frequency' given by
\begin{equation}\label{shear}
\widetilde{S} \equiv 2A - 2S\frac{k_x}{k_z} = -\Omega q_R + \Omega q_z\frac{k_x}{k_z} = \frac{\bb{k}}{k_z}\bcdot(\grad v_y)_\perp,
\end{equation}
in which $(\grad v_y)_\perp= -\Omega q_z \ex + \Omega q_R\ez$ is a vector perpendicular to the shear gradient. Thus $\widetilde{S}$ is proportional to $\delta \bb{s}\bcdot\grad v_y$ and so measures the size and sign of the shear that a fluid element experiences as it is displaced. (Note that $\widetilde{\kappa}$ and $\widetilde{S}$ are related via $\widetilde{\kappa}^2=4\Omega^2+2\Omega\widetilde{S}$.) The sign of $\widetilde{S}$ is important for both the MRI and diffusive non-ideal MHD shear instabilities. For most modes in discs $\widetilde{S}$ is negative, but when $q_z(k_x/k_z)$ is positive and sufficiently large the sign of $\widetilde{S}$ flips as it becomes dominated by the vertical shear. This has consequences for the ADSI, as we shall see, especially in Section \ref{sec:ADazimuth}. Note that $\widetilde{S}=0$ for the special wavevector orientation $k_x/k_z= q_R/q_z$.

\subsection{General dispersion relation}\label{sec:disprel}

Solvability of the linearised equations furnishes us with a fourth-order dispersion relation for the growth rate $s$. It can be written as
\begin{equation}\label{disp}
s^4+ a_3 s^3 + a_2 s^2 + a_1 s + a_0 =0,
\end{equation}
where the coefficients are
\begin{align}
a_3 &=k^2 \treta, \\
a_2 &= 2\kva^2 + \widetilde{\kappa}^2 \frac{k^2_z}{k^2} + k^4 \deteta +  k^2\eta_{xy} \widetilde{S}, \\
a_1 &=  k^2\treta \left[ \kva^2 + \widetilde{\kappa}^2
     \frac{k^2_z}{k^2} \right], \\
\refstepcounter{equation}
a_0 &= \widetilde{\kappa}^2 k^2_z k^2 \eta_{xx} \eta_{yy} +  \left[ \kva^2 - 2\Omega k^2_z \eta_{yx} + 2\Omega\widetilde{S} \frac{k^2_z}{k^2} \right]\notag \\* \mbox{} &\quad \times \left[ \kva^2 + \frac{\widetilde{\kappa}^2}{2\Omega} k^2\eta_{xy}\right] \tag{\theequation {\it a}} \label{eqn:a0a}\\*
\mbox{} &= \widetilde{\kappa}^2 k^2_z k^2 \deteta + k^2\eta_{xy}\widetilde{S}\left[ \kva^2 + \widetilde{\kappa}^2\frac{k^2_z}{k^2}\right] \notag \\* \mbox{} &\quad + \kva^2 \left[ \kva^2 + 2\Omega (k^2 \eta_{xy}-k^2_z\eta_{yx}) + 2\Omega\widetilde{S}\frac{k^2_z}{k^2}\right] . \tag{\theequation {\it b}}\label{eqn:a0b}
\end{align}
The constant term has been written in two different, yet equivalent, forms. The first, equation \eqref{eqn:a0a}, groups terms in a similar way to equation (83) of \citet{bt01} and equation (A9) of \citet{kb04}. The first of the two multiplicative bracketed factors corresponds to the radial force on a fluid element, and the second to the torque. In ideal MHD, the torque is purely Alfv\'{e}nic and serves as a restoring force: a positive angular displacement experiences a negative (magnetic) torque. In the standard MRI, the radial force is positively directed (i.e., repulsive): $\kva^2+2\Omega S (k_z/k)^2<0$. The result is that outwardly displaced fluid elements continue to move outwards. But for sufficiently large vertical shear so that $q_z(k_x/k_z) > q_R$, the radial force is attractive and the perturbations are stable -- magnetically modified epicycles. Non-ideal MHD complicates this picture by effecting a more complex interplay between the radial force and the torque: off-diagonal components of the resistivity tensor appear in both factors and can take on either sign depending on the magnetic-field and wavevector orientation. For example, if $\widetilde{\kappa}^2\eta_{xy}$ were to be negative and sufficiently large to offset the restoring magnetic tension, then the torque would be {\em repulsive}. In this situation, an {\em attractive} radial force can destabilize the disc.

The second form, equation \eqref{eqn:a0b}, groups terms in a similar way to equation (35) of \citet{kb04}, in that it separates the orbital dynamics (namely, the MRI and the interplay between the Coriolis force and the non-ideal MHD) from the coupling between the background velocity shear and the off-diagonal resistivity. Concerning the latter, note that the combination
\[
\eta_{xy}\widetilde{S} = \eta_{xy} 2A + \eta_{zy} 2S = \bb{\eta}\,\bb{:}\grad\bb{v}_0 ,
\]
where $\eta_{zy} = -(k_x/k_z)\eta_{xy}$ in axisymmetry, is the very same dyadic coupling identified by \citet{kunz08} as capable of driving diffusive shear instabilities (see Sections \ref{sec:adsi} and \ref{sec:hsi}). Off-diagonal elements in the magnetic
 diffusion tensor are essential for this.

\subsection{Dimensionless parameters}\label{sec:parameters}

The dispersion relation and its solutions depend on a large set of dimensionless numbers, the most important of which are those controlling the non-ideal MHD processes. We list these now. 

The background shear flow may be tuned by the dimensionless horizontal and vertical shear rates, $q_R$ and $q_z$. Once the wavenumbers of the modes are scaled using the vertical Alfv\'en length, $v_{{\rm A}z}/\Omega$, our incompressible disc model has no intrinsic outer scale, but at times we introduce the disc scale height $H$ to assess whether growing modes actually fit into the disc. As a result, the (vertical) plasma beta can appear in the form
\begin{equation}
    \beta_z \equiv \frac{2(\Omega H)^2}{v^2_{{\rm A}z}}.
\end{equation}

The three non-ideal effects -- Ohmic, ambipolar, and Hall -- are described respectively by the three Elsasser numbers,
\begin{align}
\Lambda_\eta \equiv \frac{v_{{\rm A}z}^2}{\kappa\eta}, \qquad  \text{Am} \equiv \frac{\nu_{\rm ni}}{\kappa}, \qquad \text{Ha} \equiv \frac{v_{{\rm A}z}^2}{v_{\rm H}^2} ,
\end{align}
where $v_{\rm H}^2 \equiv \Omega B_{0z}c/(2\pi e n_{\rm e})$ is the square of the Hall velocity introduced by \citet{bt01}. Note that $v^2_{\rm H}$ and thus ${\rm Ha}$ are defined with respect to the orbital frequency $\Omega$, while $\Lambda_\eta$ and ${\rm Am}$ make reference to the epicyclic frequency $\kappa$. This choice reflects the sensitivity of the Hall physics to rotation/shear rather than to epicylic motion. For example, if $\bb{\Omega}$ and $\bb{B}$ are oriented oppositely, $v^2_{\rm H}$ and ${\rm Ha}$ take negative values. While ambipolar diffusion is sensitive to the shear as well (through the action of the ADSI), the reference to $\kappa$ is more generally relevant to stable rotating systems, since the neutrals would undergo epicyclic oscillations if it were not for their being collisionally coupled to the flux-frozen species. Indeed, in the limit of axial fields and wavenumbers, the instability criterion for the MRI subject to Ohmic dissipation and/or ambipolar diffusion makes explicit reference to $\Lambda_\eta$ and/or ${\rm Am}$ (see equation 28 of \citealt{bb94} or equation 14 of \citealt{kb04}).

While these parameters arise naturally within an incompressible model, $\Lambda_{\eta}$ and Ha depend inconveniently on the strength of the applied field. It is useful to construct parameters independent of the field strength using $H$, namely the Ohmic magnetic Reynolds number and the Hall Lundqvist number:
\begin{align}
    \text{Rm} \equiv \frac{H^2\kappa}{\eta}, \qquad L_{\rm H} \equiv \frac{2 H v_{{\rm A}z}}{\Omega v_{\rm H}^2}  = \frac{H}{\ell_{\rm H}}.
\end{align}
Note that $L_{\rm H}$ is the ratio of the scale height to the characteristic Hall lengthscale. These quantities can be related to the earlier Elsasser numbers via
$\text{Rm}= \beta_z\Lambda_\eta\kappa^2/(2\Omega^2)$ and $L_{\rm H}= \sqrt{2\beta_z} \text{Ha}$.

\section{Analysis}\label{sec:analysis}

The dispersion relation is somewhat daunting and, in its full generality, meaningful results may only be obtained numerically. In this section we approach its analysis in distinct and separate non-ideal limits: purely Ohmic, purely ambipolar (with and without azimuthal fields), and purely Hall. This permits us to obtain definite analytic stability criteria and asymptotic growth rates that (along with numerical growth rates) fix ideas and help us construct a more complete physical picture. 
Once this is done, we then turn to more realistic PP disc models, in which all the non-ideal effects are acting simultaneously, to differing degrees (Section \ref{sec:ppdisc}). 

For each non-ideal case our approach is as follows. We are interested in regions of the disc that are \emph{MRI stable}, but possibly \emph{VSI unstable}. Our reasoning is that, if the MRI is active, then it will be the dominant player in the disc dynamics and the VSI will be unlikely to compete. The first step in each subsection then is to obtain a stability criterion for the MRI. 
The second step is to derive instability criteria for the VSI (and/or ambipolar and Hall shear instabilities) within this MRI-stable regime. We also determine its other properties (e.g., asymptotic growth rates) where possible. One difficulty with this approach is in clearly distinguishing between MRI and VSI modes -- the two live on the same branch of the general
dispersion relation \citepalias{lp18}. However, they can be disentangled on account of their preferred wavevector orientation: the MRI prefers wavevectors pointing predominantly in the vertical direction, $k_x/k_z\sim 0$, while the VSI (like the diffusive shear instabilities) prefers wavevectors pointing predominantly in the radial direction, $k_x/k_z\sim -1/q_z$. The size of the ratio $k_x/k_z$ is then our key differentiator.

Before venturing into the non-ideal cases (Sections \ref{sec:ohmic}--\ref{sec:Hall}), we recapitulate some previous results concerning the VSI, both for completeness and to set the scene. Namely, the purely hydrodynamical VSI dispersion relation (Section \ref{sec:vsi}), and the VSI and MRI in the ideal MHD limit (Section \ref{sec:idealvsimri}). We also derive the VSI growth rate and a rough stability criterion in the limit of strong general non-ideal effects (Section \ref{sec:vsi_strongeta}), which provides some guidance for the analysis presented in the subsequent subsections.

\subsection{Purely hydrodynamic VSI}\label{sec:vsi}

To remove all magnetic effects we set $\bb{v}_{\rm A}=\mathbf{0}$ in the dispersion relation
\eqref{disp}, which can then be solved straightaway for $s$. We obtain 
\begin{equation} \label{growthrate}
s^2= -\frac{k_z^2}{k^2}\widetilde{\kappa}^2.
\end{equation}
(Setting $q_R=3/2$ yields equation (13) of \citetalias{lp18}). Instability is assured whenever the effective epicyclic frequency is imaginary, $\widetilde{\kappa}^2<0$, and this occurs for perturbations with suitably oriented wavevectors:
\begin{equation}\label{hydrostab}
q_z\frac{k_x}{k_z} < -\frac{\kappa^2}{2\Omega^2} = q_R - 2,
\end{equation}
The dependence on the sign of the effective epicyclic frequency emphasises the centrifugal character of the instability. Marginal stability, $s=0$, occurs either for $\bb{k}\to \ex$ or for ${\bb{k}\parallel\grad\ell}$ \citep{ks82}, with growth being limited to wavevector orientations between these limits. Because $\grad\ell$ is almost radial, the VSI is thus restricted to a narrow arc of wavevector orientations, spanning an angle of only ${\approx\,}2 q_z\ll 1$ above the radial axis.

Because the hydrodynamical equations possess no characteristic lengthscale, the growth rate can only depend on wavevector orientation (viscosity or vertical structure will, however, introduce a lengthscale dependence; see \citetalias{lp18}). The maximum growth rate is then obtained by maximising $s$ over $k_x/k_z$. The value of $k_x/k_z$ at maximum growth is ${\approx} -\kappa^2/(q_z\Omega^2)$ for small vertical shear; hence, the maximum growth rate is given by
\begin{equation}
s_{\rm max} \approx |q_z|\frac{\Omega^2}{\kappa}.
\end{equation}
In a Keplerian disc this corresponds to $\Omega|q_z|$. Because $|q_z|\ll 1$ the growth rate can be considerably less than the orbital frequency, and the growth occurs primarily along $k_x/k_z \approx -1/q_z$ for which disturbances are radially narrow and vertically elongated.

\subsection{The limit of ideal MHD}\label{sec:idealvsimri}

Restoring $\bb{v}_{\rm A}$ in equation \eqref{disp} but setting $\bb{\eta}=0$, we obtain the follow bi-quadratic dispersion relation for the growth rate \citepalias{lp18}:
\begin{align}
 &s^4 + \left[ 2\kva^2+\widetilde{\kappa}^2\frac{k_z^2}{k^2} \right]s^2 \notag \\
 &\hskip1.5cm + \kva^2\left[
 \kva^2 +2\Omega\widetilde{S} \frac{k_z^2}{k^2}\right]=0. \label{MHDdisp}
\end{align}
The resulting instability criterion is
\begin{equation}\label{stab}
\frac{\kva^2}{\Omega^2} < -2\frac{\widetilde{S}}{\Omega} \frac{k_z^2}{k^2}.
\end{equation}
This criterion captures both the MRI and the VSI. If the effective shear rate seen by a mode, $\widetilde{S}$, can be made equal to or greater than 0, then the disc is stable. For MRI channel flows ($k_x=0$), this corresponds to the now-famous stability condition $\rmd\Omega/\rmd R>0$ (which replaces Rayleigh's criterion; \citealt{bh91}). Vertical shear slightly modifies the classical MRI problem: fastest growth occurrs at small but nonzero $k_x$, namely $k_x/k_z= -q_z/q_R$, and the maximum growth rate is $s= \sqrt{A_0^2+S^2_0}$, a factor $|q_z|$ larger than in the classical case without vertical shear. The MRI thus supplements its primary instability mechanism with that of the VSI. 

To distinguish the VSI itself, we let $k_x/k_z \sim -1/ q_z$, the wavevector orientation that maximizes VSI growth in the hydrodynamic limit. The instability criterion is then satisfied for 
\begin{equation}\label{condytion}
\kva^2\sim k^2_z v^2_{{\rm A}z}\lesssim
q_z^2\Omega^2.
\end{equation}
This additional condition does not exist in the hydrodynamical case, and brings in a lengthscale constraint on growing modes. Magnetic tension is clearly stabilising. In fact, by taking the limit $\kva\to 0$ in equation \eqref{MHDdisp} (no tension) we obtain the hydrodynamical VSI dispersion relation \eqref{growthrate}.\footnote{A subtlety here is that the limit $\kva\to 0$ is singular. For small but non-zero $\kva$, the VSI lives on the Alfv\'enic branch of the dispersion relation; for precisely $\kva=0$, the VSI lives on the inertial branch.}

Owing to the absence of an intrinsic outer scale in our incompressible disc model, it is not straightforward to interpret the constraint \eqref{condytion}. But if we concede that a mode at some vertical location $z_0$ cannot have a vertical lengthscale larger than $H$, then the instability criterion can be reframed conveniently as $\beta_z \gtrsim q_z^{-2}$ \citepalias[see][]{lp18}.

\subsection{The VSI and the limit of strong non-ideal effects}\label{sec:vsi_strongeta}

Suppose that non-ideal MHD is very strong, so that the generalised Elsasser number $\Lambda \equiv v_{\rm A}^2/(|\bb{\eta}|\Omega) \ll 1$. In this limit the $a_i$ coefficients are dominated by contributions from the resistivity tensor and, moreover, $a_3\sim a_1\sim \text{tr}(\bb{\eta})\sim \Lambda^{-1}$
and $a_2\sim a_0 \sim \text{det}(\bb{\eta})\sim \Lambda^{-2}$.
In order to extract the VSI growth rate $s$ in this regime, we assume that it is independent of the magnitude of $\bb{\eta}$ (this then excludes the MRI). Immediately we see that the leading-order balance in \eqref{disp} is ${\sim}\Lambda^{-2}$. A dispersion relation of the form $a_2 s^2+a_0=0$ results, with both $a_2$ and $a_0$ dominated by terms proportional to $\deteta$. On solving for $s$ at this order we obtain the hydrodynamical VSI growth rate, equation \eqref{growthrate}. Thus in this regime, `diffusion' by non-ideal effects (be it resistive, ambipolar or via the Hall effect) eradicates any influence of MHD on the VSI, and to leading order the stability criterion and growth rates are purely hydrodynamical.

\begin{figure*}
\center
\includegraphics{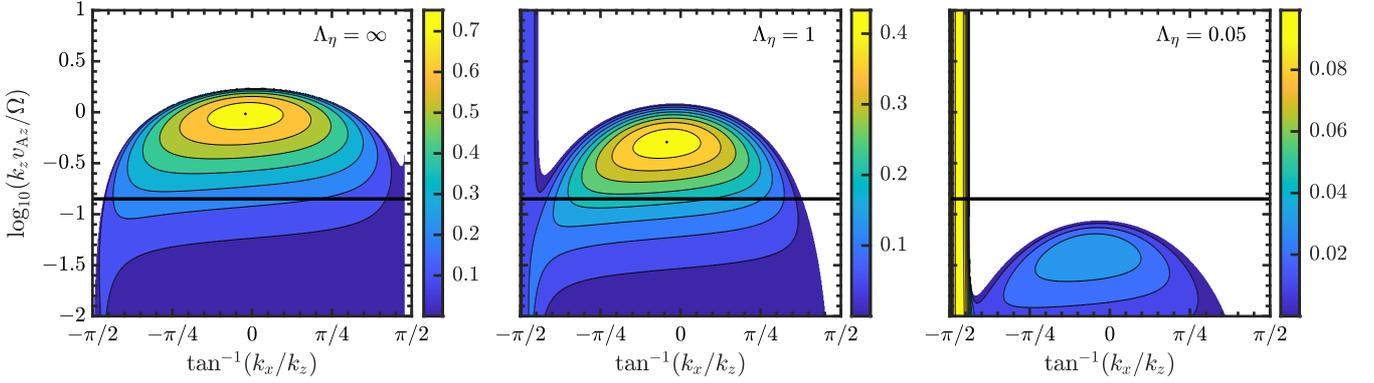}
\caption{Coloured contours of VSI and MRI growth rates in the plane of
$\tan^{-1}(k_x/k_z)$ and $k_z v_{\rm A}/\Omega$ for different Ohmic Elsasser
numbers $\Lambda_\eta$. In all cases $q_z=0.1$ and $B_y=0$. Negative growth rates and those less than $10^{-9}\Omega$ have been suppressed. The black line corresponds to the disc's vertical wavenumber when $\beta_z=100$; all modes near or below the black horizontal lines probably do not fit into the disc and thus cannot appear. The dominant MRI modes manifest around $k_x/k_z=0$ and the VSI modes around $\tan^{-1}(k_x/k_z)\approx -\tan^{-1}(1/q_z)\approx -1.471$. }
\label{fig:OhmVSI}
\end{figure*}

To make further progress we must find the next-order correction to the growth rate. This is easily obtained by writing $s= s_0 + s_1 + \dots$, where $s_0$ is the hydrodynamic VSI growth rate and $s_1\sim s_0\Lambda$. Equating terms of order $\Lambda^{-1}$ in the dispersion relation yields the first-order correction:
\begin{equation}
s_1 = \frac{1}{2}\frac{\khatva^2}{\deteta} \Biggl[\left|\frac{\widetilde{\kappa}k}{2\Omega k_z}\right| \eta_{xy}  + \left|\frac{\widetilde{\kappa}k}{2\Omega k_z}\right|^{-1}\eta_{yx} - \treta \Biggr] .
\end{equation}
How this correction scales with other parameters, in particular $q_z$, can be readily 
obtained. But in the pure Ohmic and pure ambipolar cases with $B_y=0$ it is clear that $s_1<0$ because 
$\eta_{xy}=\eta_{yx}=0$. Thus magnetic tension stabilises the VSI but its effect is partly mitigated by magnetic diffusion. In contrast, when only the Hall effect is acting the growth rate correction can be of either sign, depending on the direction of the background field and of the relative sizes of $|\widetilde{\kappa}/(2\Omega)|$ and its inverse. Then, depending on the combination $q_zk_x/k_z$, the Hall effect can either mitigate magnetic tension (as in the Ohmic case) or increase the VSI growth rate via the diffusive MRI mechanism. Something similar occurs when ambipolar diffusion is working alongside a nonzero $B_y$, though 
the algebra is far more complicated to work through.

Before moving on, note that the plasma beta does not appear in any of the criteria. This is because in the strong non-ideal limit there is no wavelength restriction on the VSI (unlike the MRI); in fact, without viscosity the VSI will operate on arbitrarily small scales (as in a hydrodynamic disc). Therefore, determining when the VSI occurs and the MRI does not requires computing (a) the critical plasma beta below or above which the MRI is suppressed, and then (b) the critical Elsasser number required for the VSI to appear on all the lengthscales unavailable to the MRI. The following subsections attempt to get a more quantitative grip on this scenario.

\subsection{Ohmic diffusion}\label{sec:ohmic}

We first examine a purely Ohmic disc, for which $\treta=2\eta$ and $\deteta=\eta^2$. Though PP discs are not governed by Ohmic diffusion alone, it nonetheless plays some role at the disc midplane at smaller radii, and offers a fairly straightforward mathematical analysis which we build upon in the later subsections.

Our approach is twofold. We first present some numerical solutions of the dispersion relation with Ohmic dissipation, and show how the MRI and VSI behave as the Ohmic Elsasser number decreases. Second, we derive some analytical estimates for when the VSI overcomes both magnetic tension and the MRI. To obtain these we first find when the MRI is suppressed by Ohmic diffusion, and then combine this criterion with one specifying when the VSI tames magnetic tension and thus extends over all wavelengths.

\subsubsection{Numerical growth rates}

In Fig.~\ref{fig:OhmVSI} we plot coloured contours of the VSI and MRI growth rates as calculated numerically from the dispersion relation \eqref{disp} for different Ohmic Elsasser numbers $\Lambda_\eta$. The thick horizontal black line corresponds to the disc's vertical wavenumber $k_z=1/H$, so that $k_z v_{{\rm A}z} /\Omega=\sqrt{2/\beta_z}$, with $\beta_z=100$. Modes beneath the black line are probably too long to fit into the disc and thus will not feature in the dynamics. Both instabilities appear on the same branch, which makes their separation difficult. However, they prefer quite different wavenumber orientation, with $\theta=\tan^{-1}(k_x/k_z)$ near $-q_z/q_R$ for the MRI and near $-\pi/2+q_z$ for the VSI. 

In the absence of Ohmic diffusion (left panel), i.e. ideal MHD, we obtain the classical MRI growth rates slightly shifted towards orientations favouring the VSI \citepalias[see discussion in][]{lp18}. The VSI is almost entirely suppressed, localised to a tiny interval of small vertical wavenumbers around $\theta=-1.471$. But as Ohmic diffusion is increased (middle and right panels) it is the MRI that is impeded, taking on smaller growth rates and relegated to smaller values of $k_z v_{{\rm A}z}$. Concurrently, the VSI asserts itself against magnetic tension, and beyond a critical value of $\Lambda_\eta\sim 1$ it can work for all values of $k_z v_{{\rm A}z}$. This is illustrated in the middle panel, though in this case the MRI is still dominant
and takes larger growth rates. For sufficiently small $\Lambda_\eta$, the MRI can be stabilised (it does not fit into the disc), as illustrated in the right panel. The VSI ,on the other hand, approaches its hydrodynamic behaviour and now completely determines the disc's stability. In the following subsections, we derive criteria for when this last situation occurs.

\subsubsection{Instability criterion}

In the dispersion relation, the coefficient controlling stability is
\[
a_0= \widetilde{\kappa}^2 k_z^2 k^2 \eta^2 + \kva^2 \left[ \kva^2 + 2\Omega\widetilde{S}\frac{k^2_z}{k^2}\right] ,
\]
with instability occurring for $a_0<0$. Assuming that $\widetilde{S}<0$, which is always the case for realistic discs and for both the VSI and MRI, the instability criterion $a_0<0$ becomes
\begin{align} \label{Ohmstab}
\frac{k_z^2 v_{{\rm A}z}^2}{\Omega^2}  < 2 \frac{(-\widetilde{S})}{\Omega} ~ \frac{f(k^2_z/k^2)}
{1+ f^2(k^2/k_z^2)(\widetilde{\kappa}^2/\kappa^2)\Lambda_\eta^{-2}},
\end{align}
where $f \equiv \left[1 - (q_z/q_R)(k_x/k_z)\right]^{-2}$. Note importantly that the inequality \eqref{Ohmstab} flips direction if the denominator in the fraction on the right side is negative; when this happens the criterion is always satisfied and instability is guaranteed for all $k_z v_{{\rm A}z}$, meaning magnetic tension drops out of the problem. This is a possibility reserved for modes that have $\widetilde{\kappa}^2<0$, such as the VSI, and will generally occur for sufficiently small $\Lambda_\eta$.

\subsubsection{The MRI}

We first determine when the MRI fails to operate in the disc. For the pure MRI the fastest-growing modes possess wavevectors $k_x/k_z \sim q_z$. As we are interested in a situation when the MRI is completely stabilised we only need examine the fastest growers.
To leading order in small $q_z$, these modes possess the instability criterion
\begin{align}\label{MRIOhm}
\frac{k^2_z v_{{\rm A}z}^2}{\Omega^2} < \frac{2q_R}{1+\Lambda_\eta^{-2}} .
\end{align}
However, they cannot have vertical wavelengths longer than the thickness of the disc, i.e.~$k_z\gtrsim 1/H$. Combining this requirement with equation \eqref{MRIOhm} yields the
rough instability criterion
\begin{equation}
\beta^2_z + 4\,{\rm Rm}^2 (1 - q_R \beta_z ) \lesssim 0.
\end{equation}
In the limit of large Rm, we keep only the terms in the parentheses and instability occurs for
subthermal
magnetic field strength: $\beta_z \gtrsim q_R^{-1}\sim 1.$\footnote{Vertically stratified models can determine this criterion exactly; for instance, if the disc is Keplerian and locally isothermal, the MRI grows when the midplane $\beta_z$ is greater than ${\approx}0.8946$. (The critical value given in \citet{Latter2010} is in error.)}
For \emph{general} Rm, inspection of the solutions using the quadratic formula reveals that a
necessary condition for the MRI to appear is $\text{Rm}\gtrsim
q_R^{-1}\sim 1$. If this is satisfied then one can show further that $\beta_z$ must lie within a range bounded by two critical values: 
the lower represents the stabilising influence of magnetic tension ($\beta_z\gtrsim 1$), whereas the upper limit represents the stabilising influence of Ohmic diffusion (and is $\beta_z\lesssim\text{Rm}^2$). Thus the MRI operates when
$1\lesssim \beta_z \lesssim \text{Rm}^2$. 

We now assume that this condition is violated -- the plasma beta is too high or too low and the MRI is not present. We then turn to the VSI to see   when it can emerge and stimulate activity in the disc.

\subsubsection{The VSI}

The fastest growing VSI modes in a hydrodynamical disc possess 
$k_x/k_z = -\kappa^2/(\Omega^2 q_z)$, but in non-ideal MHD the preferred wavevector orientation is slightly shifted, as we shall see. However, to set the scene we initially take the hydrodynamical value of $k_x/k_z$, as it clearly shows how magnetic tension and Ohmic diffusion work against each other.

When $k_x/k_z = -\kappa^2/(\Omega^2 q_z)$ we have $\widetilde{\kappa}^2=-\kappa^2$, the denominator in the criterion \eqref{Ohmstab} can be negative for sufficiently small $\Lambda_\eta$, and thus instability unrestricted.
Assuming $|q_z|\ll q_R$ and expanding everything in powers of $|q_z/q_R|$, the instability criterion becomes
\begin{align}\label{Ohmcrit}
\frac{k^2_z v_{{\rm A}z}^2}{\Omega^2} \lesssim \frac{\alpha_1\,q_z^2}{1-\alpha_2\, q_z^{-2}\Lambda_\eta^{-2}},
\end{align}
where
\[
\alpha_1 \equiv \frac{2q_R^2\Omega^4}{(4-q_R)\kappa^4}, \quad 
\alpha_2 \equiv \frac{q_R^4\kappa^4}{(4-q_R)^4\Omega^4} .
\]
Importantly, the inequality changes direction if the denominator flips sign. In a Keplerian disc, $\alpha_1=18/25$ and $\alpha_2=(3/5)^4=81/625$.

When there is no Ohmic
diffusion ($\Lambda_\eta\to\infty$) we recover the result of \citetalias{lp18}: 
$k_z v_{{\rm A}z} \lesssim q_z\Omega$. Then the VSI can only survive if
magnetic tension is very weak. However, a finite Ohmic diffusion
leads to a softening of this constraint and a larger range of $k_z$
yields instability: diffusion is working just as expected, undoing
the stabilisation issuing from the field. 
If diffusion is sufficiently
strong the denominator flips sign and there
is no restriction on $k_z v_{{\rm A}z}$: the VSI works on all vertical wavelengths. This occurs when $\Lambda_\eta < \sqrt{\alpha_2} q_z^{-1}.$ Note that the heuristic estimate given in \citetalias{lp18} is incorrect.\footnote{Their rough argument failed to set diffusion on the largest
wavenumber $k_x$. Substituting $k_x^2\eta$ for $k^2\eta$ in the
argument
there (Section 4.5) yields the above correct estimate.} Finally, because the lengthscale dependence of the problem has dropped out we need not be concerned whether the VSI can fit into the disc.

We next return to equation \eqref{Ohmstab}, let $k_x/k_z$ vary, and consider when instability is unrestricted, i.e., when the denominator is negative. For a given $k_x/k_z$, unrestricted instability is assured if
\[
\Lambda_\eta < \Lambda_{\eta,{\rm c}} \equiv \frac{\bigl(-\widetilde{\kappa}^2/\kappa^2\bigr)^{1/2}|k/k_z|}{[1- (q_z/q_R)(k_x/k_z)]^2} .
\]
As can be checked, $\Lambda_{\eta,{\rm c}}$ rises from zero at $k_x/k_z=-\kappa^2/(2q_z\Omega^2)$, takes a maximum, and then decays to zero as $q_zk_x/k_z\to -\infty$. Thus for $\Lambda_c<\Lambda_{\eta,{\rm c}}$ there is a band of $k_x/k_z$ over which the VSI is unrestricted. Straightforward calculus shows that $\Lambda_{\eta,{\rm c}}$ takes its maximum
when 
\[
\frac{k_x}{k_z}\approx  -\frac{1}{2q_z}\left(4+q_R + \sqrt{9 q_R^2-8 q_R+16}\right)
\]
to leading order in small $q_z$. This value differs by an order-unity multiplicative factor from the maximum $k_x/k_z$ of hydrodynamical growth. The corresponding maximum $\Lambda_{\eta,{\rm c}}$ is a complicated function of $q_R$, which we do not give. But for a Keplerian disc, $\Lambda_{\eta,{\rm c}} \approx 0.63856q_z^{-1}$, which is about twice the value calculated earlier for when $k_x/k_z=-\kappa^2/(q_z\Omega^2)$.

\begin{figure*}
\center
\includegraphics{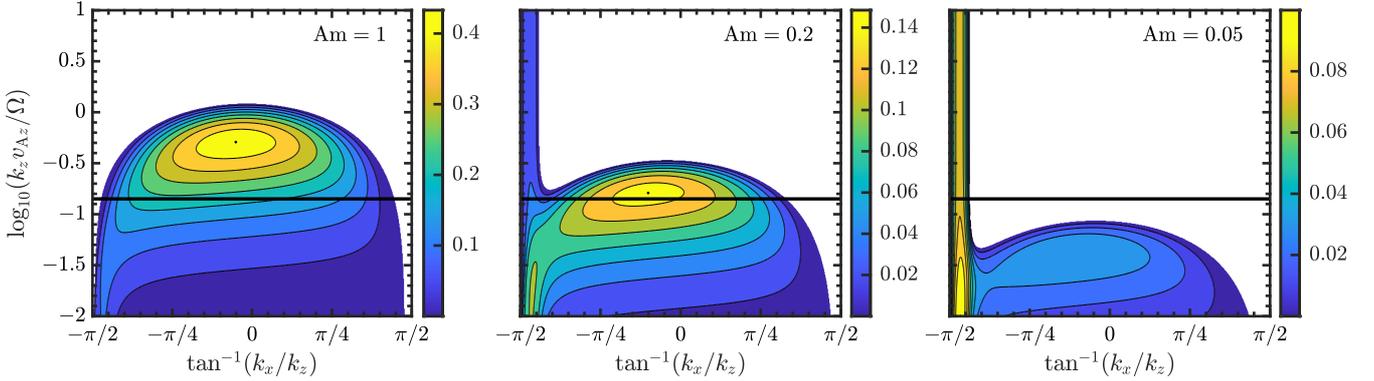}
\caption{Coloured contours of VSI and MRI growth rates in the plane of
$\tan^{-1}(k_x/k_z)$ and $k_z v_{{\rm A}z}/\Omega$ for different ${\rm Am}$. From left to right ${\rm Am}=1$, $0.2$, and $0.05$.
 As earlier, $q_z=0.1$, $B_y=0$, and the black horizontal line describes the disc's vertical wavenumber when $\beta_z=100$. }
\label{fig:ADVSI}
\end{figure*}

\subsubsection{Summary of criteria}

For the VSI to emerge and be fully active on all scales while the MRI is fully suppressed, we require $\Lambda_\eta < \Lambda_{\eta,{\rm c}}\sim q_z^{-1}$ and either $\beta_z\lesssim 1$ or $\beta_z \gtrsim \text{Rm}^2$. Recognising next that $\Lambda_\eta = 2(\text{Rm}/\beta_z)(\Omega/\kappa)^2$, we have two conditions relating $\beta_z$ and Rm. 

Let us examine the weak-field (large-$\beta$) limit first. If we assume that ${\rm Rm}>q_z$, then the stability criterion for the MRI is more difficult to satisfy than the VSI criterion. We thus conclude that the VSI dominates the disc when $\beta_z>\text{Rm}^2$. In the strong-field limit ($\beta\lesssim 1$), we find that the MRI is stabilised by magnetic tension but the VSI may still operate if Ohmic diffusion is sufficiently strong. This regime may be described by  $\text{Rm}\,q_z \lesssim \beta_z \lesssim 1$. Obviously, for very strong fields both the MRI and VSI are suppressed. In summary, in either regime we find that, when the MRI has been switched off, the VSI can usually take its place.
 
Finally, we should stress that these conditions are more stringent than in reality. For one, we have taken as the criterion for VSI dominance to be when the VSI works on all scales and the MRI is suppressed, rather than when the VSI possesses a faster growth rate. The critical limits we have put on $\beta_z$ and Rm, are thus `upper' and `lower' limits. But they have the benefit of simplicity and are, from numerical experimentation, close to the actual limits. Perhaps more importantly, it makes little sense to strive for exactness here, given the loss of accuracy inherent in pushing our local incompressible model to scales of order $H$. 
 
\subsubsection{Comparison with Cui \& Lin (2021)}

This subsection briefly compares our results to recent work on Ohmic diffusion and the VSI \citep[hereafter \citetalias{cl21}]{cl21}, which appeared while this manuscript was in preparation. The majority of the calculations in \citetalias{cl21} employed a compressible vertically stratified shearing box \citep[similar to][]{ly15} pierced by a background toroidal field, and thus conveniently precluded the MRI but unfortunately at the expense of the important effect of magnetic tension. Nonetheless, \citetalias{cl21} finds that magnetic pressure helps suppress the VSI, and that Ohmic diffusion can `undo' its effects, results that echo our calculations.
 
\citetalias{cl21} also undertakes a brief analysis in a purely local incompressible box with a net vertical background field -- an identical set-up to what we use in this subsection. The MRI and VSI growth rates are then compared as the key parameters $\beta$ and $\Lambda_\eta$ vary. \citetalias{cl21}, however, set $k_z\approx 1/H$ from the outset, which imposes a severe and artificial handicap on the MRI, especially for weaker fields (the MRI growth rates should be maximised over $k_z$).  As a consequence, the comparison is misleading and the broader conclusions reached regarding the dominance of the VSI or MRI are erroneous.

\subsection{Ambipolar diffusion with no azimuthal field}\label{sec:AD}

In this subsection we investigate the influence
of ambipolar diffusion on its own, omitting both Ohmic diffusion
and the Hall effect. In this case, $\treta=[\vasq+\khatva^2]\nu^{-1}_{\rm ni}$ and $\deteta=\vasq\khatva^2 \nu^{-2}_{\rm ni}$. To simplify the analysis, we set the
background azimuthal field to zero. This eliminates
 complications from the ADSI. We will investigate such complications separately in Section \ref{sec:ADazimuth}.

We find the mathematics and physical behaviour are very similar to the Ohmic case and thus we skip a number of steps. In particular, our numerical solutions show the same pattern as appears in Fig.~\ref{fig:OhmVSI}: when Am decreases from large to small values the MRI is pushed to long scales, ultimately longer than the disc thickness; concurrently the VSI eventually breaks free of magnetic tension and extends over all scales. We plot a sequence of numerical growth rates as Am decreases in Fig.~\ref{fig:ADVSI}. 

In the rest of this subsection we derive an analytic criterion for when the MRI is suppressed and the VSI extends over all scales. In particular, we return to the dispersion relation \eqref{disp} and examine its last term:
\begin{equation}
a_0= \kva^2\left[\kva^2+2\Omega\widetilde{S}\frac{k_z^2}{k^2} +
   k_z^2 v_{\rm A}^2 \widetilde{\kappa}^2 \nu_\text{ni}^{-2}\right].
\end{equation}
Instability occurs when $a_0<0$, a condition that can be reframed as a restriction on $k_z^2v_{{\rm A}z}^2$, as in equation \eqref{Ohmstab}.
 We now examine the MRI briefly before looking at the VSI.

\subsubsection{The MRI}

The fastest-growing MRI modes once again are channel flows
for which $k_x/k_z \to 0$. Assuming, in addition, that $|q_z|\ll 1$
the MRI instability criterion is
\begin{equation}
\frac{k_z^2 v_{{\rm A}z}^2}{\Omega^2} < \frac{2 q_R}{1+ \text{Am}^{-2}}.
\end{equation}
 Note that
we have used $v_{\rm A} \approx v_{{\rm A}z}$, true to leading order in $q_z^2$. 
Evidently, as $\text{Am}$ decreases and AD starts to dominate, the
MRI is banished to ever longer vertical wavelengths. In fact, no MRI modes
fit in the disc once $\beta_z  \lesssim q_R^{-1}(1+ \text{Am}^{-2})$. 
Clearly, this condition reduces to the ideal MHD case when $\text{Am}\to\infty$,
but differs appreciably from the ideal criterion when $\text{Am}<1$.
Unless ambipolar diffusion is dominant, only very strong magnetic fields stabilise the MRI and, in contrast to the Ohmic MRI, there is no stabilisation when the field is very weak.

\subsubsection{The VSI}

The analysis for VSI modes is very similar to the Ohmic case. To set the scene we home in first on the fastest-growing hydrodynamical VSI mode and 
set $k_x/k_z= -\kappa^2/(\Omega^2 q_z)$, for small $q_z$. Then to leading order
the instability criterion is
\begin{align}\label{ADcrit}
\frac{k_z^2 v_{{\rm A}z}^2}{\Omega^2} < \frac{\alpha_1 q_z^2}{1- \alpha_3\text{Am}^{-2}},
\end{align}
where $\alpha_3\equiv q_R^2/(4-q_R)^2$. In a Keplerian
disc, $\alpha_3= (3/5)^2$. Note that the inequality changes direction if the denominator on the right side changes sign.
Just as in the Ohmic case,
ambipolar diffusion counteracts magnetic tension and permits the
VSI to grow, ultimately, on all lengthscales. 
The latter occurs when the denominator in equation \eqref{ADcrit} is less
than 0. This happens when
$\text{Am} < \alpha_3^{1/2}$
which importantly does not involve $q_z$, to leading order in small $q_z$.

The orientation of fastest hydrodynamic growth, $k_x/k_z= -\kappa^2/(\Omega^2 q_z)$, though a useful starting point does not generally give the dominant VSI mode in non-ideal MHD, as we now show. For general $k_x/k_z$, the criterion for unrestricted VSI on all vertical scales is ${\rm Am}<{\rm Am}_{\rm c}$, where
\[
{\rm Am}_{\rm c} \approx \frac{\bigl(-\widetilde{\kappa}^2/\kappa^2\bigr)^{1/2}}{[1- (q_z/q_R)(k_x/k_z)]^2},
\]
to leading order in small $q_z$. This critical Am has a maximum value when $k_x/k_z=-(\kappa^2/\Omega^2+q_R)q_z^{-1}$ for which ${\rm Am}_{\rm c} \approx q_R/(2\sqrt{4-2q_R})\sim 1$. In a Keplerian disc, ${\rm Am}_{\rm c}\approx 3/4$, which is slightly larger than $\alpha^{1/2}=3/5$.

In summary, for the MRI to be switched off we require $\beta\lesssim (1+{\rm Am}^{-2})$
and for the VSI to concurrently extend over all scales, we require that ${\rm Am}<{\rm Am}_{\rm c}\sim 1$. 
This situation is favoured, thus, by stronger fields. Though it should be pointed out that when $\text{Am}\gtrsim 1$ both instabilities may suppressed when the magnetic
field is too strong.

\begin{figure*}
\center
\includegraphics{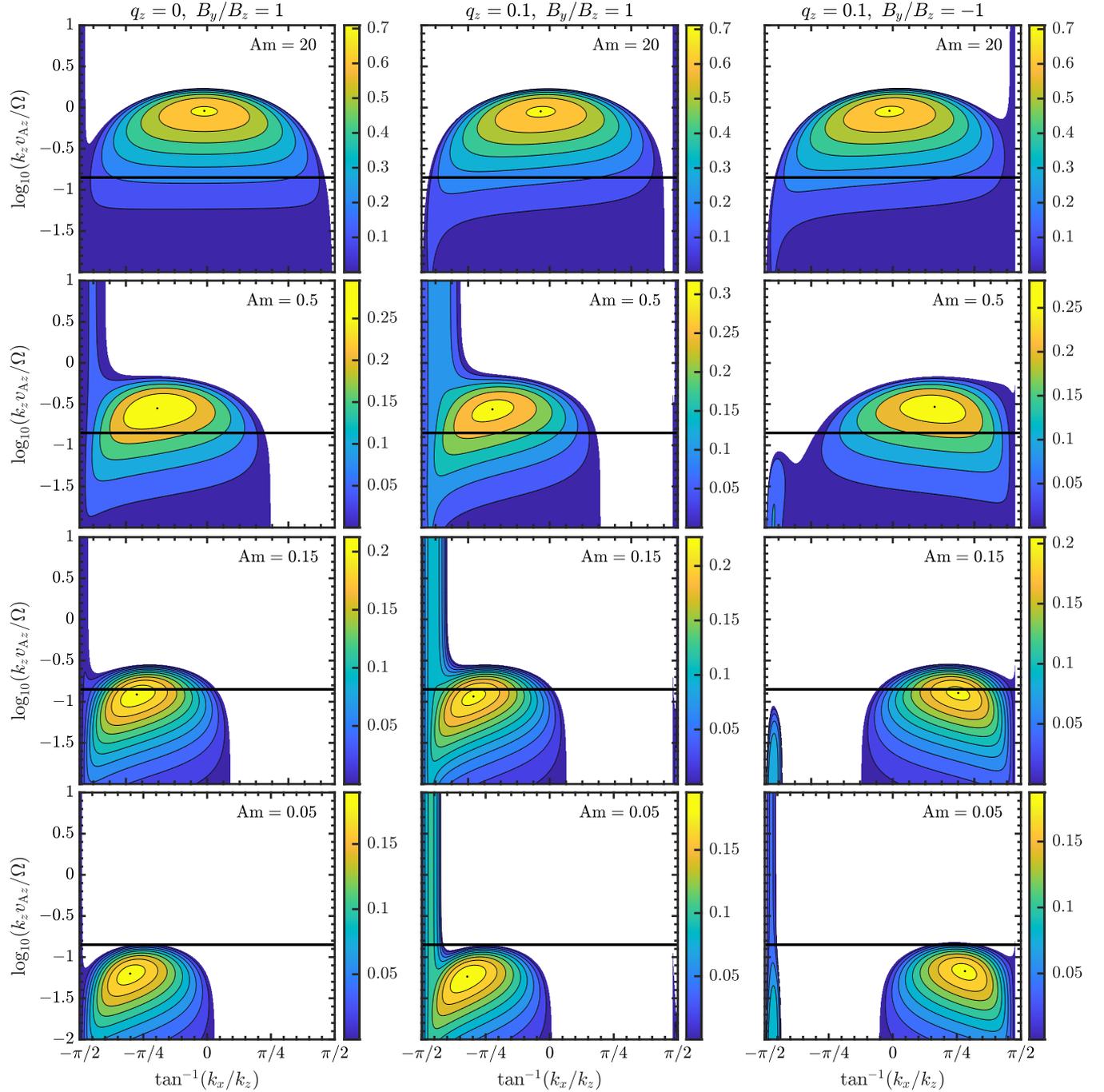}
\caption{Coloured contours of growth rates in the plane of $\tan^{-1}(k_x/k_z)$
and $k_z v_{{\rm A}z}/\Omega$ for different ${\rm Am}$,
with and without vertical shear, and for different signs of $B_y/B_z$. The black line corresponds to the disc's vertical wavenumber when $\beta_z=100$.}
\label{fig:ADVSI2}
\end{figure*}

\subsection{Ambipolar diffusion with azimuthal fields}\label{sec:ADazimuth}

When an azimuthal field is combined with ambipolar diffusion the stability properties of the
gas change remarkably. Even in the absence of vertical shear, weak growth
extends to very large $|k_x/k_z|$; this may be regarded as a separate
instability mechanism to the MRI, relying on ambipolar diffusion and
shear -- the ADSI \citep{kb04,desch04,kunz08}. 
The ADSI and the VSI can potentially share the same general wavevector orientation if $q_z$ and $B_y/B_z$
possess the same sign, and we find that if this is the case the two modes combine into one hybrid instability. It is also possible for the ADSI to occupy two separate $k_x/k_z$ bands.

As earlier, we first present some illustrative numerical solutions to the dispersion relation, and then derive stability criteria in different cases to help elucidate these results and their underlying physics. We examine separately the cases of no rotation, rotation but no vertical shear, and then the full problem, from which we establish criteria for when one or the other instability dominates in the disc.

\subsubsection{Numerical growth rates}

Fig.~\ref{fig:ADVSI2} shows contours of growth rates in the $\tan^{-1}{(k_x/k_z)}$--$k_z v_{{\rm A}z}/\Omega$ plane with and without vertical shear and for different values of
$\text{Am}$, namely $20$, $0.5$, $0.15$, and $0.05$. As noted, the left column corresponds to a disc with no vertical shear ($q_z=0$), while the other two columns have $q_z=0.1$. In the first two columns $B_{y}=B_{z}$, and in the last column $B_{y}=-B_{z}$.

For the parameters in the left column, the VSI is inactive because $q_z=0$. But in addition to the MRI modes (localised initially around $k_x/k_z\sim 0$)
there is a slower-growing band of ADSI modes on all scales for sufficiently large and negative $k_x/k_z$. These additional modes favour an intermediate range of $\text{Am}$: when ${\rm Am}$ is either large or small, the instability band is extremely narrow and grows very slowly. 

The middle column in Fig.~\ref{fig:ADVSI2} reinstates vertical
shear while retaining $B_y=B_z$. This appears to have several effects. If we examine large negative $k_x/k_z$, the vertical shear strengthens and widens the band of ADSI for intermediate ${\rm Am}$. In fact, we view this mode as a hybrid that extracts energy from
both the vertical and radial shears and thus takes some of its character from both the VSI and ADSI. In the bottom-most panel (${\rm Am}=0.05$), however, the unstable
mode may be unambiguously assigned to the VSI. Next, if we examine large positive $k_x/k_z$, we see that at larger values of $\text{Am}$ there appears a separate narrow band of weak instability. We attribute this branch to an additional ADSI that works solely on vertical shear. It appears only for $k_x/k_z>q_R/q_z$ and thus when the effective shear rate $\widetilde{S}$ is positive and dominated by $q_z$. For our choice of parameters, this corresponds to $\tan^{-1}(k_x/k_z)>1.471$. For smaller $\text{Am}$ the growth rate of this branch is exceptionally small.

Finally, in the right column we flip the sign of $B_y/B_z$. This means that the wavevector orientations favoured by the VSI and the principle ADSI separate out. We can now observe the three different instabilities as we vary $\text{Am}$. The top panel shows the MRI and the ADSI, with the latter now oriented so that its preferred wavevector orientation has $k_x/k_z>0$. In the central panels, at smaller $\text{Am}$, we witness the slow emergence of the VSI at large negative $k_x/k_z$ and long wavelengths (small $k_z v_{{\rm A}z}/\Omega$). In the bottom-most panel, at even smaller $\text{Am}$, the MRI and ADSI are suppressed while the VSI now extends over all scales. In this case, there appears to be no secondary ADSI working solely on the vertical shear.

\subsubsection{General instability criterion}

When $B_y\neq 0$ the formula for $a_0$ is rather involved. But the instability criterion $a_0<0$ may be written as
\begin{align} \label{ADfull}
\frac{k_z^2v_{{\rm A}z}^2}{\Omega^2} <
\frac{2f(-\widetilde{S}/\Omega)(k_z^2/k^2)\left[1-g(B_y/B_z)\widetilde{\kappa}^2/(2\nu_\text{ni}\Omega)\right]}
{1+f(B/B_z)^2(\widetilde{\kappa}/\nu_\text{ni})^2 - g(B_y/B_z)(\widetilde{S}/\nu_\text{ni})},
\end{align}
where $g\equiv f(k_x/k_z + q_z/q_R)$. As before, when the denominator is less than 0 the inequality flips direction.
This instability condition governs the MRI, the ADSI, and
the VSI.
In the absence of vertical shear ($q_z=0$) the criterion simplifies to equation (37) 
in \citet{kb04}. When $B_y=0$, it agrees with equation \eqref{ADcrit} in Section \ref{sec:AD}. Though it is complicated, informative limits can be extracted.

\subsubsection{The MRI}

Once again, we distinguish the MRI from the other two instabilities via its preferred
wavevector orientation, $k_x/k_z \approx 0$. Thus we take $f\approx 1$ and $g\approx 0$.
If $\widetilde{S}<0$ and $\widetilde{\kappa}>0$ (true for wavevector orientations favouring the MRI), then instability proceeds on scales sufficiently long, with the denominator never changing sign. However, for small $\nu_\text{ni}$ the denominator ultimately becomes large due
to the positive term proportional to $\nu_\text{ni}^{-2}$ and the range of unstable scales is banished to unfeasibly long lengthscales. 
The resulting instability criterion is similar in form 
to Section \ref{sec:AD} except that the azimuthal field helps stabilise the mode. Instability occurs when 
\begin{equation}
\beta_z > q_R^{-1}(1+ \text{Am}^{-2}B^2/B_z^2).
\end{equation}
For azimuthal fields significantly stronger than vertical fields, the MRI is killed off for a larger range of $\beta_z$; specifically subthermal fields stabilise the MRI even when $\text{Am}>1$. It is this stabilisation mechanism that undergirds the laminar saturated MRI states witnessed in several recent simulations \citep[e.g.][]{bs13,lkf14,gressel15}.

\subsubsection{Shear but no rotation}

We now turn to the large-$|k_x/k_z|$ instabilities -- the ADSI and VSI -- starting with the former in its purest non-rotating form.
In the absence of rotation, $\widetilde{\kappa}^2/(2\Omega)\to \widetilde{S}$ and $\Omega\to 0$ otherwise. We also set $q_z=0$ and thus retain radial shear only, for simplicity. The instability criterion becomes
\begin{align}
\frac{B_y}{B_z}\frac{k_x}{k_z}\widetilde{S}> \nu_\text{ni},  
\end{align}
which can be satisfied on all lengthscales, even for large $\nu_\text{ni}$, provided that $(k_x/k_z)(B_y/B_z)\widetilde{S}>0$. This limit corresponds to the ADSI explicated in \citet{kunz08}; see equation (21) in that work. To make better comparison with the next subsection, the instability criterion can be reworked into
\begin{equation}\label{norot}
\frac{|\widetilde{S}|}{\kappa}\left|\frac{k_x}{k_z}\right|> \text{Am}\left|\frac{B_z}{B_y}\right|.  
\end{equation}

\subsubsection{Rotation but no vertical shear}\label{sec:ADSI}

If next we add in rotation but assume no vertical shear, so that $\widetilde{\kappa}^2=\kappa^2>0$,
we find that the ADSI is somewhat stabilised. To simplify the discussion, we determine only when the mode is free of magnetic tension and thus extends over all scales, i.e., when the denominator in equation \eqref{ADfull} is negative. This is only possible for wavevector orientations satisfying $(k_x/k_z)(B_y/B_z)\widetilde{S}>0$, as above; though because $q_z=0$, $\widetilde{S}=-q_R\Omega<0$, and thus the requirement simplifies to $(k_x/k_z)(B_y/B_z)<0$. If this holds, we obtain extended 
instability when
\begin{align} \label{ADSI}
\frac{|\widetilde{S}|}{\kappa}\left| \frac{k_x}{k_z} \right| > \text{Am}\left|\frac{B_z}{B_y} \right|+\text{Am}^{-1}\left(\left|\frac{B_z}{B_y}\right|+\left|\frac{B_y}{B_z}\right|\right) .
\end{align}
So for a given ${\rm Am}$ we can always find a wavevector orientation that yields
extended instability. Note, however, the new second term on the right-hand side of the inequality when comparing with equation \eqref{norot}: instability is now curtailed in the limit of both large and small ${\rm Am}$ \citep{kb04}, in agreement with the numerical solutions. Intermediate ${\rm Am}$ is the most propitious for extended ADSI
mode growth. In addition, as Am becomes small, the growth rate appears to scale as $\text{Am}^2$, and thus quickly becomes subdominant to other processes. 

\subsubsection{Rotation, vertical shear, and $q_z(B_y/B_z)<0$: pure VSI}

On reinstating the vertical shear, the problem gets rather intricate. Vertical shear permits $\widetilde{S}$ and $\widetilde{\kappa}^2$ to take both positive and negative values [depending on $q_z(k_x/k_z)$],
and we see immediately this changes the balance of power in the denominator
of the instability criterion. In particular, the term proportional to
$\nu_\text{ni}^{-2}$ can move from stabilising to destabilising, and it is this term that can dominate in 
in the limit of small $\nu_\text{ni}$. Furthermore, the $k_x/k_z$ dependence of the effective shear $\widetilde{S}$ permits up to two separate bands of the ADSI, one influenced more by the vertical shear, and the other by the orbital shear. Finally, it is possible for the VSI and ADSI to combine, by occupying the same range of $k_x/k_z$.
Both instabilities favour large $|k_x/k_z|$, with the VSI requiring $(k_x/k_z)q_z <0$ and the ADSI requiring $(k_x/k_z)(B_y/B_z)\widetilde{S}>0$. Thus when $q_z(B_y/B_z)>0$ both inequalities are satisfied and the two combine into a single hybrid mode. When $q_z(B_y/B_z)<0$ they separate out and inhabit bands of $k_x/k_z$ of different sign. In this subsection we deal with the latter case, as it is simpler, and then treat other cases in following subsections. 

We examine modes for which $q_z(k_x/k_z)<0$ and are thus potentially VSI unstable. Next, in order to establish when the VSI extends over all scales free of magnetic tension, we ask when the denominator in the criterion \eqref{ADfull} is negative. This occurs, for a given $k_x/k_z$, when Am is sufficiently small:
\begin{equation*}
\text{Am} < \text{Am}_{1,{\rm c}} \approx  - q_z\left(\frac{B_z}{B_y}+\frac{B_y}{B_z}\right)\frac{\Omega(\kappa^2/\Omega^2+ 2 q_z k_x/k_z)}{\kappa(q_R-q_z k_x/k_z)q_z(k_x/k_z)},
\end{equation*}
true to leading order in small $q_z$. The critical Am rises from 0 at $ (k_x/k_z)=-\kappa^2/(2\Omega^2 q_z)$ and then decays to 0 for large negative $k_x/k_z$, as in previous sections. It takes its maximum value at $k_x/k_z= q_z^{-1}\left(q_R-2 - \sqrt{4 - 2 q_R}\right)$, to leading order in small $q_z$, at which point Am$_{1,{\rm c}}$ equals
\begin{align}
\text{Am}_\text{max} &\approx -2q_z\left(\frac{B_z}{B_y}+\frac{B_y}{B_z}\right) \nonumber\\*
\mbox{} &\times \frac{1}{(2+\sqrt{4-2q_R})(2+\sqrt{4-2 q_R}-q_R)}.
\end{align}
Evidently, the VSI is far less favoured when $B_y\neq 0$. For small $q_z$, we have now that $\text{Am} \lesssim |q_z|$, rather than $\text{Am}\lesssim 1$, as in Section \ref{sec:AD} with $B-y=0$. Note that our asymptotic analysis is based on $B_y/B_z$ not taking small values, and in reality as $B_y/B_z\to 0$ we should have Am$_\text{max}$ approach an order 1 finite value (cf. Section \ref{sec:AD}) rather than diverge.

\subsubsection{Rotation, vertical shear, and $q_z(B_y/B_z)>0$: the hybrid VSI/ADSI}

When $q_z(B_y/B_z)>0$ and for wavevectors $q_z k_x/k_z<0$ we capture both the VSI and ADSI. Both favour large values of $|k_x/k_z|$ and in their purest form extend over all scales. As earlier, we derive the condition when the denominator in equation \eqref{ADfull} is negative. This occurs when
\begin{equation*}
\text{Am}<\text{Am}_{2,{\rm c}} \approx\frac{1}{q_z} \left(\frac{B_y}{B_z}\right)\frac{\Omega q_R^2(-q_z k_x/k_z)}{\kappa(q_R-q_z k_x/k_z)}  
\end{equation*}
to leading order in small $q_z$. Unlike Am$_{1,{\rm c}}$, this critical value has no turning point, but rises from 0 (near $k_x/k_z=0$) and asymptotes to a maximum value as $q_z k_x/k_z\to -\infty$. This value is 
\begin{equation}
\text{Am}_\text{max} \approx \left(\frac{\Omega}{\kappa}\right)\left(\frac{B_y}{B_z}\right) \left(\frac{q_R^2}{q_z}\right).
\end{equation}
As a consequence, the VSI's preference for $k_x/k_z\sim -1/q_z$ does not appear in the onset of instability, but does come in through the relative sizes of growth rates. As ${\rm Am}$ is decreased, modes with the largest $|k_x/k_z|$ are destabilised first (which we associate with the ADSI mechanism) but their growth remains small as ${\rm Am}$ decreases further. Growth instead appears to be maximised near the VSI orientation, $k_x/k_z\sim -1/q_z$. The critical ${\rm Am}$ for onset is much larger than in the previous subsection; ${\rm Am}_\text{max}$ here is ${\sim}1/q_z$ rather than ${\sim}q_z$. This reinforces the idea that for ${\rm Am}$ near marginal instability the onset is governed by the ADSI, not the VSI.

On the other hand, the ADSI manifesting here differs from its cousin with no vertical shear (Section \ref{sec:ADSI}), where for any ${\rm Am}$ one can always find an unstable mode of some (potentially extreme) $k_x/k_z$. This is not the case when vertical shear is added: above ${\rm Am}_{2,{\rm c}}$ no instability is possible. We attribute this to the non-constant effective shear, which is wavevector dependent. As $|k_x/k_z|$ increases, evidently, the effective shear becomes less conducive to instability.

\subsubsection{Rotation and vertical shear: pure ADSI}

We now characterise the ADSI uncontaminated by the VSI. Thus we examine wavevector orientations $q_z(k_x/k_z)>0$, and treat the cases of $q_z(B_y/B_z)$ positive and negative separately (as above). Key to this discussion is the critical orientation $(k_x/k_z) = q_R/q_z$, as it corresponds to when $\widetilde{S}$ passes from negative to positive values. Recalling that the ADSI works only when $\widetilde{S}(k_x/k_z)(B_y/B_z)>0$, if $q_z(B_y/B_z)<0$ instability proceeds when $k_x/k_z< q_R/q_z$, and if $q_z(B_y/B_z)>0$ when $k_x/k_z> q_R/q_z$. Note that the latter band of ADSI is in addition to its manifestation in the hybrid mode for $q_z k_x/k_z<0$.  

We zoom in on the critical wavevector orientation and expand the stability criterion around that point. 
After some algebra we find the criterion to leading order is
\begin{equation}\label{asymptotic}
    \frac{k_z^2 v_{{\rm A}z}^2}{\Omega^2} <  q_R \text{Am}\left(\frac{ q_zB_y B_z}{B^2}\right)\left(q_z\frac{k_x}{k_z}-q_R \right)^{-1},
\end{equation}
where we have also taken $q_z\ll q_R$. When $q_z B_y/B_z<0$ the right-hand side is positive for $k_x/k_z< q_R/q_z$ and diverges as $k_x/k_z\to q_R/q_z$ from below, thus extending instability to all scales; the right-hand side is negative, however, when $k_x/k_z> q_R/q_z$ and instability is impossible for these wavevector orientations. When $q_z B_y/B_z>0$ the exact opposite is the case. The two instability bands on either side of $\tan^{-1}(k_x/k_z)= 1.504$ in the top panels of the middle and right columns of Fig.~\ref{fig:ADVSI2} illustrate this behaviour.

\subsubsection{Summary}

Though rather complicated, it is possible to boil down these results into a compact set of criteria for when (a) the MRI is stabilised
and (b) the VSI or hybrid VSI/ADSI takes over. 
To satisfy the former, the magnetic field needs to be
sufficiently strong so that $\beta \lesssim 1+ \text{Am}^{-2}(1+B_y^2/B_z^2)$. 
To then decide on whether the VSI breaks free of tension we must know the local orientation of $B_y/B_z$ and the vertical shear $q_z$: (a) for negligible or small $B_y/B_z$, we require ${\rm Am}\lesssim 1$; (b) for non-negligible $B_y/B_z$ and if $q_z B_y/B_z<0$, then we need ${\rm Am}\lesssim |q_z B_y/B_z|$; and (c) if $q_z B_y/B_z<0$, then ${\rm Am}\lesssim |q_z^{-1}B_y/B_z|$, though in this case the VSI and ADSI have merged into the same mode. 
In principle we should also account for the $q_zk_x/k_z>0$ branch of the ADSI, though numerically we find its growth rate to be smaller than the VSI modes, certainly when ${\rm Am}$ is small. We hence will omit it in our discussion.

\subsection{The Hall effect}\label{sec:Hall}

In this final subsection we omit ambipolar and Ohmic diffusion, and turn to the Hall effect. In this case, $\treta=0$ and $\deteta=[c\hat{\bb{k}}\bcdot\bb{B}/(4\pi en_{\rm e})]^2=\khatva^2\ell^2_{\rm H}$. 
As shown in \citet{wardle99} and \citet{bt01}, the MRI undergoes quite radical changes in the presence of the Hall effect, depending on the polarity of the vertical magnetic field relative to the rotation vector. The key parameter here is $\text{Ha}$, defined in Section \ref{sec:parameters}. Here we explore what happens when vertical shear is added to the mix.

\subsubsection{Numerical solutions}

\begin{figure*}
\center
\includegraphics{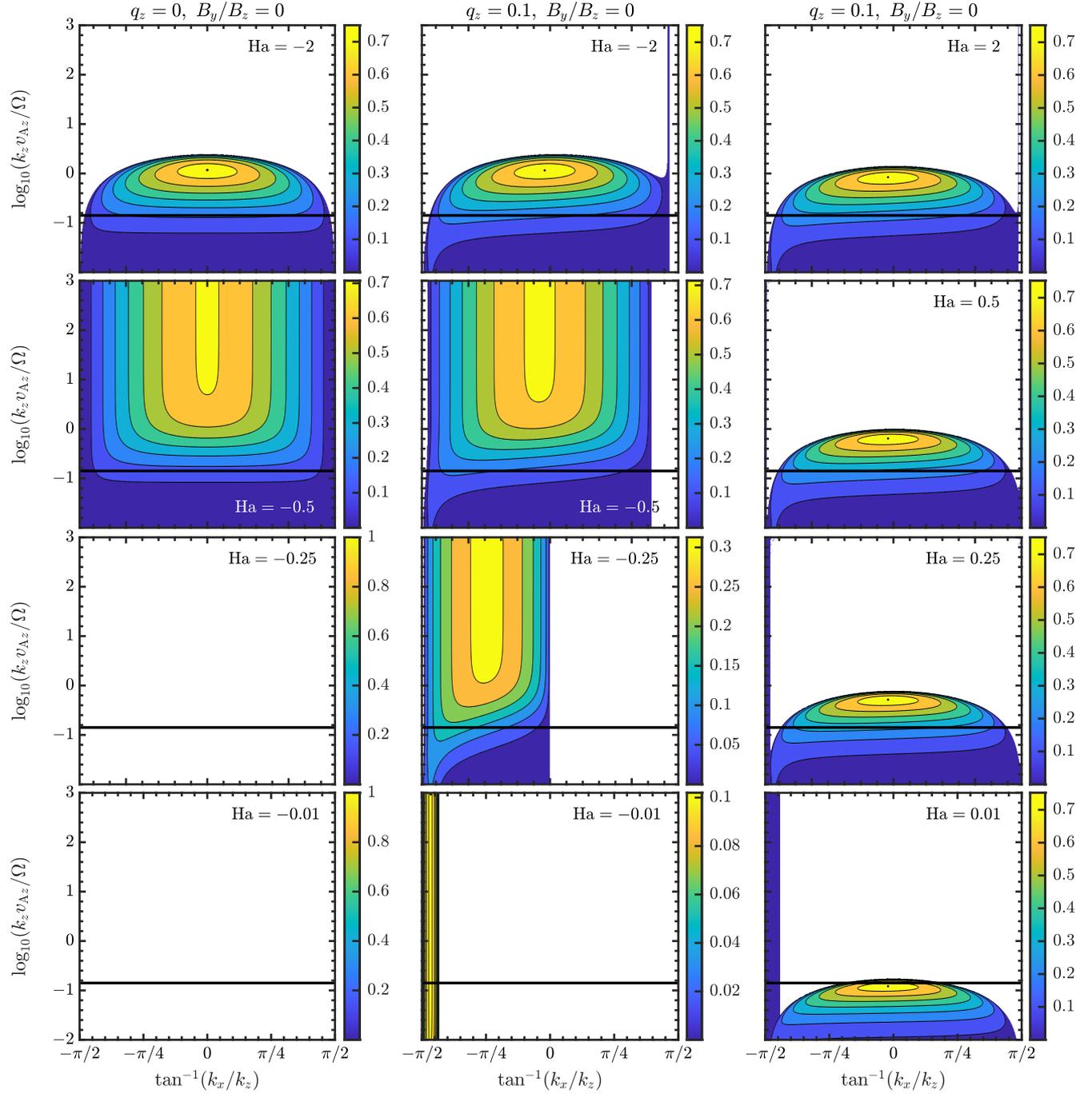}
\caption{Coloured contours of growth rates in the plane of $\tan^{-1}(k_x/k_z)$ and $k_z v_{{\rm A}z}/\Omega$ for different ${\rm Ha}$ and with or without vertical shear. In the left column $q_z=0$, otherwise $q_z=0.1$. The black line corresponds to the disc's vertical wavenumber when $\beta_z=100$.}
\label{fig:HVSI}
\end{figure*}

In Fig.~\ref{fig:HVSI} we plot growth rates as functions of $\tan^{-1}(k_x/k_z)$ and $k_z v_{{\rm A}z}/\Omega$ for different ${\rm Ha}$ and with and without vertical shear. 
In the first column we sample several negative ${\rm Ha}$ but set $q_z=0$. Beginning from the top, $\text{Ha}=-2$ and we recover the MRI. For smaller negative values the range of instability explodes rather abruptly; this is the DMRI, and occurs when $-{\rm Ha}$ drops below $1$. In the panel second from the top we set ${\rm Ha}=-0.5$ and instability extends over all wavevector orientations and on all scales, though it favours $k_x/k_z\sim 0$. At small enough values of $-{\rm Ha}$ (e.g., $0 > {\rm Ha}> -1/4$ for a Keplerian disc if $k=k_z$), instability shuts down entirely, just as abruptly as the DMRI switches on. The lower two panels in the first column correspond to ${\rm Ha}=-0.25$ and $-0.01$; clearly there is no growth. We omit calculations using positive ${\rm Ha}$ and $q_z=0$.

In the middle column we let ${\rm Ha}$ take the same negative values as in the left column but now with vertical shear turned on, $q_z=0.1$. The most important difference here to the $q_z=0$ case is that instability never abruptly cuts off as $-{\rm Ha}$ takes smaller values. Instead, we witness a smooth morphing of the DMRI into the VSI, with instability migrating from all wavevector orientations to those focused on $k_x/k_z\sim -1/q_z$, and a reduction in the growth rate from ${\sim}|A|$ to ${\sim}|S|$. The other difference is an additional narrow band of weak instability at large positive $k_x/k_z$, evident in the top panel. This we attribute to a secondary HSI reliant solely on the vertical shear.

In the right column, we employ the same sequence of \emph{positive} ${\rm Ha}$ as in the other columns, setting $q_z=0.1$. The picture here appears to be more in line with the Ohmic and ambipolar cases with no azimuthal field: as ${\rm Ha}$ decreases the unstable modes are gradually pushed to longer scales, and once ${\rm Ha}\lesssim 1$ the VSI emerges and attacks all scales with wavevectors $k_x/k_z\sim-1/q_z$.  For ${\rm Ha}\lesssim 1$ the unstable modes with $k_x/k_z\approx 0$ are a blend of the MRI and the HSI, and in the limit ${\rm Ha}\ll 1$ can be attributed mostly to the HSI: then it is predominantly the right-handedness of the whistler wave, rather than the radially directed magnetic tension force, that generates $\delta B_x$ at the expense of $\delta B_y$ and thereby completes the feedback loop with the Keplerian shear.

\subsubsection{General stability criteria}

Having sketched out numerically the various behaviours brought out by the vertical shear and the Hall effect, we derive the analytical conditions governing the onset of instability. This is made somewhat easier by the fact that the purely Hall dispersion relation is bi-quadratic, with a constant term $a_0$ that may be factored as a product of the two coefficients (see equation \eqref{eqn:a0a})
\begin{align*}
b_1 &= \kva^2 + \kva k_z v_{{\rm A}z} \text{Ha}^{-1} + 2\Omega\widetilde{S}\frac{k_z^2}{k^2} , \\
b_2 &= \kva^2 + \frac{\widetilde{\kappa}^2}{4\Omega^2} \kva k_z v_{{\rm A}z} {\rm Ha}^{-1}.
\end{align*}
In all previous cases, determining the sign of $a_0$ was sufficient to establish instability or not, but that is not necessarily the case with Hall and vertical shear, and we also need to check the sign of $a_2$, which is given by
\begin{equation*}
a_2= 2\kva^2 +  \widetilde{\kappa}^2\frac{k_z^2}{k^2} + \frac{k^2}{4\Omega^2 k_z^2}\kva^2 k_z^2 v_{{\rm A}z}^2 {\rm Ha}^{-2}.
\end{equation*}
Stability requires both $a_2$ and $a_0$ to be positive. In fact, we find that $a_2>0$ for most parameter values, only flipping sign on the longest lengthscales (and for $q_z\neq 0$). In what follows, $a_2$ only features in our discussion of positive Ha, and instability is usually assured when $a_0<0$
(thus $b_1$ and $b_2$ must differ in sign). Finally, without loss of generality, let us assume that $k_z v_{{\rm A}z}>0$.

\subsubsection{No vertical shear}\label{sec:Hall_qz0}

We first check what happens when there is no vertical shear ($q_z=0$). We start with the (more interesting) case of $\text{Ha}<0$. As ${\rm Ha}$ takes smaller and smaller negative values we progress through three regimes. Regime 1 corresponds to sufficiently large and negative ${\rm Ha}$, i.e.~$\text{Ha}<-1$, upon which instability occurs for
\begin{align} \label{Hally}
\frac{k^2_z v_{{\rm A}z}^2}{\Omega^2} < \frac{2q_R (k_z^2/k^2)}{1+\text{Ha}^{-1}}.
\end{align}
This we associate with the standard MRI, slightly modified by the Hall effect (top-left panel in Fig.~\ref{fig:HVSI} with ${\rm Ha}=-2$). Regime 2 corresponds to  $-1 < \text{Ha} < -\kappa^2/(4\Omega^2)$, an interval upon which instability occurs with no restrictions on the value of $k_z$. We associate this regime with the DMRI (second panel in the left column of Fig.~\ref{fig:HVSI} with ${\rm Ha}=-0.5$). Finally, regime 3 corresponds to small negative ${\rm Ha}$, viz.~$0> \text{Ha} > -\kappa^2/(4\Omega^2)$, for which no instability is possible (bottom two panels in the left column of Fig.~\ref{fig:HVSI}). 

When $\text{Ha}>0$ we have a combination of the MRI and HSI, which on decreasing $\text{Ha}$ is pushed to longer and longer scales. In this case, equation \eqref{Hally} can be manipulated into a condition involving the plasma beta and the Hall Lundqvist number; we find that the disc is completely stabilised by the Hall effect when $\beta \lesssim 1+ \text{Ha}^{-1}$, which can be re-framed as
\begin{equation} \label{HallMRI}
\beta \lesssim \frac{1}{4}\left(\sqrt{2}L_{\rm H}^{-1} + \sqrt{2 L_{\rm H}^{-2}+4} \right)^2.
\end{equation}
In the ideal limit $L_{\rm H}\gg 1$ this criterion returns to $\beta \lesssim 1$, and in the Hall-dominated limit $L_{\rm H}\ll 1$, we obtain $\beta \lesssim 2 L_{\rm H}^{-2}$.

\subsubsection{Vertical shear and $\text{Ha}<0$}

We next restore the vertical shear and assume $\text{Ha}<0$. As shown numerically, it is not straightforward to disentangle the primary DMRI and the VSI. Assuming $a_2>0$, and thus neglecting very long scales, then setting $b_2>0$ and $b_1<0$, we find that instability attacks all other scales for wavevectors oriented in the range
\begin{equation}\label{gg}
    (1+\text{Ha}^{-1})q_R < q_z\frac{k_x}{k_z} < \frac{q_R(\text{Ha}^{-1} \kappa^2/\Omega^2+4)}{2(2-q_R\text{Ha}^{-1})}.
\end{equation}
For all negative $\text{Ha}$ this range always exists, and thus instability always appears. It thus revises regime 3 in Section \ref{sec:Hall_qz0}, in which no instability is possible. 
The above criterion covers unstable modes on all scales, but instability also extends to the parameter region $q_z (k_x/k_z)<(1+\text{Ha}^{-1})q_R$ on sufficiently long scales:
\begin{equation}
\frac{k_z^2 v_{{\rm A}z}^2}{\Omega^2} < \frac{k_z^2}{k^2}\frac{2q_R^2}{1-(q_z/q_R)(k_x/k_z)+\text{Ha}^{-1}}.    
\end{equation}

We now examine two limits. If we let $\text{Ha}$ approach $0$ from below, then the lower bound on the range of instability in equation \eqref{gg} recedes to negative infinity, while the right bound turns negative at $\text{Ha}=-\kappa^2/(4\Omega^2)$, and then approaches the finite value $-\kappa^2/(2\Omega^2)$. When $\text{Ha}\to 0^{-}$, i.e.~the limit of strong Hall, the instability condition becomes simply $\widetilde{\kappa}<0$, the hydrodynamical criterion for the VSI. We can see this migration of the instability band to increasingly negative $k_x/k_z$ rather clearly in the middle column of Fig.~\ref{fig:HVSI}

On the other hand, when $\text{Ha}$ approaches $-\infty$, the range of $k_x/k_z$ allowing unconditional instability narrows to a vanishingly small interval around $k_x/k_z=q_R/q_z$ (the condition for $\widetilde{S}=0$). This very narrow band of instability we associate with the secondary vertical HSI. It can be best observed in the top panel of the middle column in Fig.~\ref{fig:HVSI}. 

\subsubsection{Vertical shear and $\text{Ha}>0$}

Finally, we examine the case of positive ${\rm Ha}$. According to our numerical solutions, both the MRI/HSI and VSI feature separately, which simplifies the analysis somewhat. 
The MRI and HSI, favouring modes with
$k_x/k_z$ small, act identically to the case with no vertical shear (to leading order in small $q_z$), with its onset controlled by equation \eqref{Hally}. Thus the criterion \eqref{HallMRI} continues to hold.

Turning next to the VSI, as in previous sections we first examine the fastest growing hydrodynamic mode and set $k_x/k_z = -\kappa^2/(\Omega^2 q_z)$. We want to determine when these modes grow on almost all scales. After some manipulation, we find that setting $a_0<0$ requires $\text{Ha}$ to be smaller than a critical number
\begin{equation} \label{false}
{\rm Ha} < {\rm Ha}\equiv\frac{\kappa^2}{4\Omega^2} \frac{q_R}{4-q_R} ,
\end{equation}
and for perturbations to lie in the following range of scales
\begin{equation} \label{true}
\frac{k^2_z v^2_{{\rm A}z}}{\Omega^2} > \frac{2\Omega^4}{\kappa^4} \frac{q^2_R q^2_z}{4-q_R + q_R {\rm Ha}^{-1}}.
\end{equation}
On the other hand, it is straightforward, though tedious, to show that condition \eqref{true} also corresponds to $a_2>0$. It follows that instability is controlled solely by criterion \eqref{false}. There is no condition on wavenumber, for we have $a_2 a_0 <0$ whether \eqref{true} is satisfied or not.

We emphasize that the instability criterion \eqref{false} only governs VSI modes with $k_x/k_z=-\kappa^2/(\Omega^2 q_z)$ and that the restriction can be less onerous for modes that have other $k_x/k_z$, especially those for which $k_x/k_z\to -\infty$ (though these modes grow at a substantially smaller rate). It is possible to find the critical ${\rm Ha}$ below which \emph{all} VSI modes extend to arbitrarily small scales. We first recognise that this is controlled by the sign of $b_2$, thus set $b_2=0$, and then solve for ${\rm Ha}$ as a function of negative $q_z k_x/k_z$. Maximising this expression, we find that the critical ${\rm Ha}$ is simply $q_R/2$. The difference between this value and that appearing in equation \eqref{false} is an order-unity factor, and will be unimportant in practice because the condition for no MRI/HSI, equation \eqref{HallMRI}, will be more difficult to satisfy. In any case, the critical ${\rm Ha}$ below which the VSI grows unconditionally is in accord with our numerical solutions shown in the right column of Fig.~\ref{fig:HVSI}.

\subsubsection{Summary}\label{sec:Hallsummary}

We summarise our results here for Hall-MHD. When $\text{Ha}>0$ things are somewhat simpler. The MRI/HSI is switched off for sufficiently small $\beta$, a criterion described by equation \eqref{HallMRI}, and the fastest-growing VSI works freely on all scales when $\text{Ha}< q_R/2$. We expect the first criterion to be the more difficult to satisfy. 

When $\text{Ha}<0$, it is less easy to be so definitive on account of the potential merging of the DMRI and the VSI. Recognising that when $\text{Ha}> -\kappa^2/(4\Omega^2)$ there is no instability in the absence of vertical shear, we label as VSI the growing mode that appears in this otherwise stable regime. In terms of the plasma beta and the Hall Lundqvist number, the criterion is hence $\beta > 16(\Omega/\kappa)^4L_{\rm H}^2.$

\subsection{Combined criteria for VSI dominance}
\label{CriteriaSummary}

We now collate all the various criteria derived in the previous subsections and combine them into (a) a general condition for MRI suppression/saturation, and (b) a condition for the VSI to free itself of magnetic tension and thus operate unhindered in the disc. Unfortunately, there are several cases, especially associated with the sign of Ha and $B_y/B_z$, but the overall set of criteria is manageable, as we shall see when applied in Section \ref{sec:ppdisc}.

\subsubsection{Suppression/saturation of the MRI}

We separate the analysis into positive and negative Ha, i.e. if the net vertical field threading the disc is aligned or anti-aligned with the rotation vector, beginning with ${\rm Ha}>0$, for which there is no DMRI. It only takes one of the three non-ideal MHD effects to halt MRI growth, and thus we take the union of the various conditions for MRI suppression in the previous subsections, rather than their intersection. We stipulate that the MRI is suppressed if any one of the following criteria are satisfied:
\begin{equation} \label{MRIkill}
\beta \gtrsim \text{Rm}^2, \qquad
\beta \lesssim \text{max}\left(1,\, \text{Am}^{-2}B^2/B_z^2,\, 2 L_{\rm H}^{-2} \right).
\end{equation}

For ${\rm Ha}<0$, i.e.~antialigned rotation and vertical field, things are more complicated because of the possibility of the DMRI, and moreover because the DMRI and the VSI can merge for intermediate ${\rm Ha}$. Section~\ref{sec:Hallsummary} presents a rough criterion for when the instability has more of the character of the VSI than the DMRI, and we will use that here. Also, to keep things simple, we leave off the effects of Ohmic and ambipolar diffusion on the onset of the DMRI. In summary, for negative ${\rm Ha}$, we stipulate that the MRI and DMRI have morphed into a form entirely or mostly exhibiting the character of the VSI when $\beta \gtrsim 16 L_{\rm H}^2$.

\subsubsection{Unfettered VSI}
   
We next assume that either (a) it is impossible for the MRI (or DMRI) to grow on account of the high diffusivities in the disc, or (b) the MRI/DMRI has grown but has then saturated in the form of a laminar magnetic structure, via the combination of $B_y$ growth and ambipolar diffusion (and also possibly by the production of a magnetic wind). Each of these cases provides a relatively quiescent environment for the VSI to grow out of. The criteria in the previous subsection certainly cover case (a), but it might also give some rough idea for when case (b) occurs.
   
Whichever, the transition between the situation when the VSI is restricted to very long scales by magnetic tension and the situation when it extends over almost all scales is abrupt and relatively easy to determine. In general, the transition occurs before the MRI is suppressed, as we increase non-ideal effects. However, things are more complicated when Ha is negative (because of the merging of the VSI and DMRI) and when $B_y/B_z>0$ (because of the merging of VSI and the ADSI). We deal with these cases separately.
 
When ${\rm Ha}>0$ and $q_z B_y/B_z<0$, the satisfaction of any one of the following criteria leads to unrestricted VSI: 
\begin{align} \label{VSIlive}
     &\beta \gtrsim \text{min}\left( q_z\text{Rm},\, 40 L_{\rm H}^2\right), \\ &\text{Am} \lesssim \text{min}\left[\left|q_z \left(\frac{B_y}{B_z}+ \frac{B_z}{B_y}\right)\right|,\,1\right].
\end{align}
When $q_z B_y/B_z>0$, the condition on ${\rm Am}$ should be replaced by ${\rm Am}< \text{max}(|q_z^{-1}B_y/B_z|,\,1)$, though one should recognise that the ensuing instability may have the character of both the VSI and ADSI.
 
For negative ${\rm Ha}$ we impose no extra condition, as the emergence of the VSI out of the DMRI is covered by the criterion shown in the previous subsection.

\section{Application to a PP disc model}\label{sec:ppdisc}

In this section we construct a global 2D disc model, specifying its density, temperature, and ionisation fraction as functions of $R$ and $z$. Maps of the three important dimensionless numbers, ${\rm Rm}$, ${\rm Am}$, and $L_{\rm H}$, follow, allowing us to apply the various criteria derived in the previous section at selected locations, and thus assess the prevalence or not of the VSI, at least within the many assumptions of the model. 

We must at the outset emphasise the large uncertainties and approximations we make here. Discs exhibit a wide range of large-scale density profiles \citep[e.g.][]{isella2009,guill2011,tazzari2017}, which can differ markedly from the one we adopt, and moreover are often punctuated by abrupt structures, such as rings, gaps and asymmetries that will complicate any analysis. Next are the poorly constrained parameters that bear on the disc's ionisation: the cosmic-ray flux, and the size  and spatial distribution of dust grains. Finally, there is the relatively unknown magnetic equilibrium state of the disc. Our results will depend sensitively on all these uncertain ingredients, and thus should be interpreted as only suggestive. The procedure, nonetheless, hopefully encourages more comprehensive future analyses.

\subsection{Disc model and key parameters}

\subsubsection{Thermodynamic structure}

For simplicity, we obtain surface density $\Sigma$ and midplane temperature $T_0$ profiles by adopting a MMSN (\citealt{hayashi81}), which in cgs units gives us
\begin{equation}
\Sigma = 1700 R_{\rm au}^{-3/2}, \qquad T_0= 280 R_{\rm au}^{-1/2},
\end{equation}
where $R_{\rm au}$ is cylindrical radius in ${\rm au}$. We next approximate the disc as locally isothermal, and moreover that the density is determined from $\rho=\rho_0\,\exp[-z^2/(2H^2)]$, with
\begin{equation}
    \rho_0= 1.4\times 10^{-9} R_{\rm au}^{-11/4}
\end{equation}
being its midplane value, assuming a gas composed of 80\% molecular hydrogen and 20\% helium, orbiting a solar-mass star. It follows that $H/R=0.042 R_{\rm au}^{1/4}$.

In reality, power-law discs such as the above exhibit vertical shear and possess density profiles slightly different to the Gaussian form enforced above; but for our purposes the discrepancies are unimportant. We do assume, however, vertical shear to be present and  in what follows estimate its magnitude roughly as $q_z\approx H/R \sim 0.1$ \citep{nelson13,bl15}.

\subsubsection{Ionisation structure}

Our method for calculating the magnetic diffusivities uses the chemical model and generalised Ohm's law prescribed in Section 4 and Appendix B of \citet{km09}, similar to what is done in \citet{lkf14} and \citet{simon15}.

We assume that the gas is mainly ionised by cosmic rays at a rate $\zeta_\text{cr}=10^{-17}\exp[-\Sigma'(z)/(96 \text{g cm}^{-2})]~{\rm s}^{-1}$, where $\Sigma'(z)$ is the $z$-dependent column mass density integrated from the top of the disc towards the midplane \citep{un80}; we acknowledge the great uncertainty in the flux of such radiation \citep{un09}. X-ray ionisation from 3~keV photons issuing from the protostar is included \citep[see][]{bg09}, as is the radioactive decay of short-lived radionuclides at a rate $10^{-19}~{\rm s}^{-1}$.       

Dust is assumed to be of a single size, but three different sizes are chosen for the calculations: 0.3, 1, and 10 microns. We do not model very small grains such as PAHs \citep{bai11b}, and caution that if a significant number are present our diffusivities will be much greater.
Six locations in the disc are selected and the relevant density and temperature input into the model. The six locations correspond to radii of 2, 10, and 50~au, at two different heights each: $0.5H$ and $2H$. We avoid the midplane itself as there the vertical shear should be precisely zero.

\begin{figure}
\center
\includegraphics{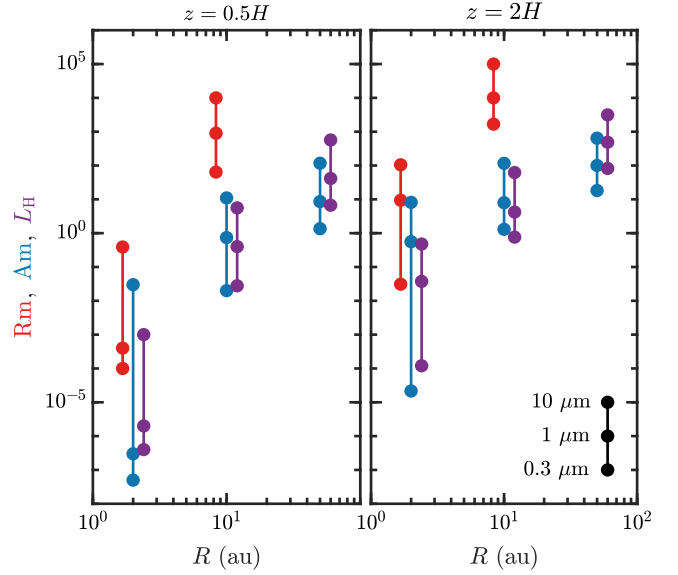}
\caption{Ranges of the magnetic Reynolds number (${\rm Rm}$, red), ambipolar Elsasser number (${\rm Am}$, blue), and Hall Lundqvist number ($L_{\rm H}$, purple) at three disc radii $R=2$, 10, and 50~au, and two different vertical locations $z=0.5H$ and $2H$. Three different dust grain sizes are modelled -- 0.3, 1, and 10 microns -- corresponding to the lower, middle, and upper dots, respectively, in the plot.}
\label{fig:ionisation}
\end{figure}

Fig.~\ref{fig:ionisation} plots the Ohmic magnetic Reynolds numbers (red), ambipolar Elsasser numbers (blue), and Hall Lundqvist numbers (purple) for the three radial locations and three dust sizes at $z=0.5H$ (left panel) and at $z=2H$ (right panel).

\subsubsection{Magnetic equilibrium and parameters}
\label{Mageqm}

In our local model the global magnetic configuration manifests in two independent parameters. The first describes the strength of the vertical field, the second the strength of the azimuthal field. The radial field is determined from these and the rate of vertical shear (cf.~Section \ref{sec:equil}). 

We keep the analysis manageable by selecting only one value for the  vertical plasma beta near the midplane, setting $\beta_z=10^3$ at $z=0.5H$. This then means that at $z=2H$ the plasma beta will drop by about an order of magnitude (keeping the net vertical field constant). Thus we take $\beta_z=100$ at the upper locations we consider. Justification for this choice of vertical field strength can be found in recent simulations of protostellar core collapse \citep[e.g.,][]{masson16,xk21b}, and the direct measurement of vertical fields in TW Hya \citep{vlemmings19}.
 
The azimuthal field strength, expressed through the parameter $B_y/B_z$ cannot easily be detached from the question of MRI stabilisation/saturation. If the disc is truly magnetically inactive, perhaps we may not expect $B_y$ to differ significantly in magnitude from $B_z$. But if conditions are such that the MRI (or DMRI) is permitted to grow, simulations and direct calculations indicate that it can saturate in a laminar state by increasing $B_y/B_z$ to a suitably large value (${\gtrsim}10$) and/or forming an outflow \citep[e.g.][]{salmeron11,bs13,lkf14,bl20}. 
 We thus allow $B_y/B_z$ to take values up to of order 10; we exclude greater values because the total plasma beta would be unrealistically low.
If the MRI is only stabilised when $B_y/B_z$ is much greater than 10, we conclude that laminar magnetic states are unavailable to the disc, and the MRI saturates by forming turbulence. We assume further that the VSI will be unable to compete with this turbulence.

\subsection{Stability boundaries at $R= 2~{\rm au}$}

We begin our survey in the inner disc at a radius of 2~au, and first set ${\rm Ha}>0$. 
Noting that ${\rm Rm}<10$ for almost all grain sizes at both heights $z=0.5H$ and $2H$, equation \eqref{MRIkill} indicates that the MRI/HSI is stabilised by Ohmic diffusion. The only exception is when we take the largest grains $a=10\,\mu{\rm m}$ and when $z=2H$; in this special case the MRI may be active in these upper layers.
Given that the MRI/HSI is absent for almost all parameter values selected, we turn to see if the VSI is free of magnetic tension, and we find that condition \eqref{VSIlive} is always satisfied thanks again to Ohmic resistivity. 

Next we treat ${\rm Ha}<0$. For all vertical heights and grain sizes we have $16L^2_{\rm H}=3.6$ at most, and usually it is much less. Therefore the condition for the suppression of the MRI/DMRI and its replacement by the VSI is always satisfied by several orders of magnitude (i.e., we require $\beta> 16L^2_{\rm H}$).

In summary, we expect the MRI/HSI/DMRI to be suppressed and the VSI to be active in the inner disc. The only exception is when there are no grains or very large grains (and then only for ${\rm Ha}>0$ and at $z>2H$).

\subsection{Stability boundaries at $R=10~{\rm au}$}

At intermediate radii, ${\sim}10~{\rm au}$, the situation becomes more complicated and very much dependent on precise parameter values. As a consequence, we discuss the two vertical heights separately.

\subsubsection{At $z=0.5H$}

In the main body of the disk the MRI cannot be stabilised by Ohmic diffusion whatever the grain size (${\rm Rm}>60$). We hence turn to the Hall effect and ambipolar diffusion, setting ${\rm Ha}>0$ first. For small particles, $a=0.3\,\mu{\rm m}$, we have $L_{\rm H}=0.028$ and the MRI/HSI modes do not fit into the disc when $\beta_z=10^3$, as assumed;  this is, however, not the case for grains of micron size or larger. It is possible that when $a=1\,\mu{\rm m}$ the ambipolar diffusion saturates the MRI in a laminar state (${\rm Am}=0.74$), but only if $B/B_z\gtrsim 30$, according to criterion \eqref{MRIkill}. Such a field configuration appears to be marginally possible given extant calculations of disc equilibria (cf.~Section~\ref{Mageqm}). For $a=10\,\mu{\rm m}$, we have ${\rm Am}=11$ and require unfeasibly large azimuthal fields, such that the total plasma beta is of order or less than 1; we thus assume that the MRI/HSI breaks down into turbulence, and the VSI cannot feature. Next, if we restrict ourselves to particle sizes of $1~\mu{\rm m}$ or less, and thus to MRI/HSI stability, we find that the VSI works on all scales free of tension thanks to Ohmic diffusion; cf.~criterion \eqref{VSIlive} with ${\rm Rm}<900$ and $q_z\sim 0.1$.

We now turn to negative ${\rm Ha}$. The criterion for the emergence of the VSI from the DMRI is once again that $\beta> 16L_{\rm H}^2$. This is just about satisfied for all grain sizes if we assume that $\beta=10^3$, since $L_{\rm H}<5.6$. For stronger vertical fields this will not be the case, however, and the VSI will be dominated by the MRI/DMRI mode.

In summary, we might expect the VSI to be operational in PP discs at ${\sim}10~{\rm au}$ and in the main body of the disk, but only if the gas is filled with sub-micron-sized grains. 

\subsubsection{At $z=2H$}

Higher up in the disc the situation gets increasingly difficult for the VSI. 
According to our model, Ohmic diffusion and the Hall effect when ${\rm Ha}>0$ are too weak to stabilise the MRI for our chosen vertical field strength (we have ${\rm Rm}>1600$ and $L_{\rm H}>0.76$). This leaves only ambipolar diffusion. 

Taking ${\rm Ha}>0$ first, we find that grain sizes of $a=0.3\,\mu{\rm m}$ permit the laminar saturation of the MRI/HSI if $B/B_z\gtrsim 16$ as ${\rm Am}=1.3$, assuming $\beta_z=100$ at this height. This is, in fact, a reasonable ratio for a laminar magnetic equilibria. But larger grain sizes require unrealistic values, $|B/B_z|> 80$, and thus we expect the MRI/HSI to break down into turbulence, which will suppress the VSI.
Howwever, even at $a=0.3\mu{\rm m}$ and $B_y/B_z<0$ it seems marginal whether the VSI can be liberated from magnetic tension, according to our criteria. And, if $B_y/B_z>0$, the VSI merges with the ADSI.  

When ${\rm Ha}<0$, only for small particles ($a=0.3\,\mu{\rm m}$) does the VSI mechanism triumph over the DMRI (at least for $\beta_z=100$); otherwise the DMRI dominates.

In conclusion, at 10~au and $z=2H$, the VSI will struggle to emerge independently. Only for particle sizes ${\sim}0.1\,\mu{\rm m}$ or smaller is it possible, but even then, when Ha$>0$ and $B_y/B_z>0$, it will combine into a hybrid ADSI--VSI mode.

\subsection{Stability boundaries at $R=50~{\rm au}$}

The situation in the outer disc is somewhat simpler, at least according to our chosen disc model. If ${\rm Ha}>0$, then at both $z=0.5H$ and $2H$ and for our three grain sizes the MRI is easily working and probably instigating turbulence -- its laminar saturation requires a $B_y$ that is too strong. We do not expect the VSI to be able to compete with the MRI in this case.

For negative ${\rm Ha}$, at $z=0.5H$ and for 0.3-micron grains then we have that the DMRI merges with the VSI when $\beta_z=10^3$, but only just so (as $L_{\rm H}=6.7$). At $z=2H$, however, $L_{\rm H}$ is too large and the DMRI has little to no character of the VSI, no matter the grain size.

\section{Conclusion}\label{sec:conclusion}

Our aim in this paper has been to explore the opposing influences of magnetic tension and non-ideal MHD on the VSI in PP discs. As has been shown elsewhere, magnetic tension can easily stabilise the VSI, banishing it to unrealistically long length scales \citepalias{lp18}. However, non-ideal effects, such as ambipolar and Ohmic diffusion and the Hall effect, can undermine the tension force and thus potentially reinvigorate the instability. The linear problem combining all the relevant physics -- vertical shear, a net magnetic field, the three non-ideal terms -- is exceptionally complicated, not only algebraically but also physically, because of the emergence of new diffusive shear instabilities that compete alongside the MRI. To make some headway we adopt the simplest possible model, the local incompressible shearing box, in which the background equilibrium is straightforward, but cooling and buoyancy are omitted. We leave to future work the vertically stratified problem, and the challenging tasks of calculating appropriate magnetic equilibria to perturb \citep[e.g.,][]{salmeron11,lo19,bl20} and of including the correct cooling physics \citep{ly15,pk21,fukuhara14}.

After deriving the governing dispersion relation for the system, we calculate growth rates numerically and obtain various analytical stability criteria for the various modes. Because of the problem's algebraic difficulties, we make several strong assumptions when obtaining these criteria: we consider the VSI to be important only when (a) the MRI is either dead or has saturated (via ambipolar diffusion) in a laminar steady equilibrium state, and (b) the VSI has overcome magnetic tension and can work on all length scales (as it does hydrodynamically). A more precise comparison of the VSI and MRI would be founded on a comparison of their maximum growth rates, but the algebra is too involved to derive clean results. As such, our criteria are probably stronger than in reality, but have the advantage of being simple and easy to apply. A separate physical difficulty is that the VSI can merge with the ADSI or the DMRI, and potentially lose its identity. In these cases, we define the VSI to be `present' when its preferred hydrodynamical wavevector orientation exhibits unrestricted growth. The combined criteria for `VSI emergence', involving all non-ideal effects, are stated in Section \ref{CriteriaSummary} by equations \eqref{MRIkill} and \eqref{VSIlive}. These are perhaps the main achievements of the paper, and it is our hope that these criteria can help researchers assess the prevalence of the VSI in their model disc equilibria or diagnose and plan MHD simulations in which the VSI may appear.

As a demonstration, we apply the criteria to an illustrative (though not necessarily representative) PP disc model based on the MMSN, allowing for some parameters to be tuned, most importantly the dust grain size. In summary, we found that at 2~au the VSI should appear unproblematically for most grain sizes considered, both in the main body of the disc and in its upper layers. On the other hand, at 50~au the VSI will struggle to get working, mainly because the MRI/DMRI or the HSI will produce turbulence that should overwhelm it. Radii around 10~au are marginal, as one might expect; within about one scale height, the VSI will probably emerge when dust particles are less than 1~micron in size, but in the upper layers the VSI is disfavoured. We, of course, caution that these claims are not necessarily true for other disc models; certainly circumstances will be much better for the VSI if there is a preponderance of smaller grains (e.g., PAHs; \citealt{bai11b}), which significantly reduce the ionisation fraction (see simulations of \citealt{cb20}, who adopt a constant ${\rm Am}\lesssim 1$ profile). On the other hand, younger Class 0 and I discs will likely sustain gravitoturbulence, which should outcompete the slower and less vigorous VSI \citep{kratter16}. Further predictions made via the application of our framework await improved disc modelling, particular concerning the dust size distribution and the strength and geometry of magnetic fields in PP discs.

\section*{Acknowledgments}

This work was started during HNL's brief sabbatical visit to Princeton in 2019. It is a pleasure to thank Richard Nelson, Can Cui, and Min-Kai Lin for useful discussions; and especially Geoffroy Lesur and Steve Balbus, our referee, for productive comments on the text.

\section*{Data Availability}

The data underlying this article will be shared on reasonable request to the corresponding author.

\bsp	
\label{lastpage}

\begin{thebibliography}{}
\makeatletter
\relax
\def\mn@urlcharsother{\let\do\@makeother \do\$\do\&\do\#\do\^\do\_\do\%\do\~}
\def\mn@doi{\begingroup\mn@urlcharsother \@ifnextchar [ {\mn@doi@}
  {\mn@doi@[]}}
\def\mn@doi@[#1]#2{\def\@tempa{#1}\ifx\@tempa\@empty \href
  {http://dx.doi.org/#2} {doi:#2}\else \href {http://dx.doi.org/#2} {#1}\fi
  \endgroup}
\def\mn@eprint#1#2{\mn@eprint@#1:#2::\@nil}
\def\mn@eprint@arXiv#1{\href {http://arxiv.org/abs/#1} {{\tt arXiv:#1}}}
\def\mn@eprint@dblp#1{\href {http://dblp.uni-trier.de/rec/bibtex/#1.xml}
  {dblp:#1}}
\def\mn@eprint@#1:#2:#3:#4\@nil{\def\@tempa {#1}\def\@tempb {#2}\def\@tempc
  {#3}\ifx \@tempc \@empty \let \@tempc \@tempb \let \@tempb \@tempa \fi \ifx
  \@tempb \@empty \def\@tempb {arXiv}\fi \@ifundefined
  {mn@eprint@\@tempb}{\@tempb:\@tempc}{\expandafter \expandafter \csname
  mn@eprint@\@tempb\endcsname \expandafter{\@tempc}}}

\bibitem[\protect\citeauthoryear{{Bai}}{{Bai}}{2011}]{bai11b}
{Bai} X.-N.,  2011, \mn@doi [\apj] {10.1088/0004-637X/739/1/51}, \href
  {https://ui.adsabs.harvard.edu/abs/2011ApJ...739...51B} {739, 51}

\bibitem[\protect\citeauthoryear{{Bai}}{{Bai}}{2014}]{bai14}
{Bai} X.-N.,  2014, \mn@doi [\apj] {10.1088/0004-637X/791/2/137}, \href
  {https://ui.adsabs.harvard.edu/abs/2014ApJ...791..137B} {791, 137}

\bibitem[\protect\citeauthoryear{{Bai} \& {Goodman}}{{Bai} \&
  {Goodman}}{2009}]{bg09}
{Bai} X.-N.,  {Goodman} J.,  2009, \mn@doi [\apj]
  {10.1088/0004-637X/701/1/737}, \href
  {https://ui.adsabs.harvard.edu/abs/2009ApJ...701..737B} {701, 737}

\bibitem[\protect\citeauthoryear{{Bai} \& {Stone}}{{Bai} \&
  {Stone}}{2013}]{bs13}
{Bai} X.-N.,  {Stone} J.~M.,  2013, \mn@doi [\apj]
  {10.1088/0004-637X/769/1/76}, \href
  {https://ui.adsabs.harvard.edu/abs/2013ApJ...769...76B} {769, 76}

\bibitem[\protect\citeauthoryear{{Balbus}}{{Balbus}}{2003}]{balbus03}
{Balbus} S.~A.,  2003, \mn@doi [\araa] {10.1146/annurev.astro.41.081401.155207}, \href
  {https://ui.adsabs.harvard.edu/abs/2003ARA&A..41..555B} {41, 555}

\bibitem[\protect\citeauthoryear{{Balbus}}{{Balbus}}{2011}]{balbus11}
{Balbus} S.~A.,  2011, in Garcia P.~J.~V., ed., {Physical Processes in Circumstellar Disks around Young Stars}. Univ.~Chicago Press, Chicago, p.~237

\bibitem[\protect\citeauthoryear{{Balbus} \& {Hawley}}{{Balbus} \&
  {Hawley}}{1991}]{bh91}
{Balbus} S.~A.,  {Hawley} J.~F.,  1991, \mn@doi [\apj] {10.1086/170270}, \href
  {https://ui.adsabs.harvard.edu/abs/1991ApJ...376..214B} {376, 214}

\bibitem[\protect\citeauthoryear{{Balbus} \& {Terquem}}{{Balbus} \&
  {Terquem}}{2001}]{bt01}
{Balbus} S.~A.,  {Terquem} C.,  2001, \mn@doi [\apj] {10.1086/320452}, \href
  {https://ui.adsabs.harvard.edu/abs/2001ApJ...552..235B} {552, 235}

\bibitem[\protect\citeauthoryear{{Barker} \& {Latter}}{{Barker} \&
  {Latter}}{2015}]{bl15}
{Barker} A.~J.,  {Latter} H.~N.,  2015, \mn@doi [\mnras]
  {10.1093/mnras/stv640}, \href
  {https://ui.adsabs.harvard.edu/abs/2015MNRAS.450...21B} {450, 21}

\bibitem[\protect\citeauthoryear{{B{\'e}thune} \& {Latter}}{{B{\'e}thune} \&
  {Latter}}{2020}]{bl20}
{B{\'e}thune} W.,  {Latter} H.,  2020, \mn@doi [\mnras]
  {10.1093/mnras/staa908}, \href
  {https://ui.adsabs.harvard.edu/abs/2020MNRAS.494.6103B} {494, 6103}

\bibitem[\protect\citeauthoryear{{B{\'e}thune}, {Lesur}  \&
  {Ferreira}}{{B{\'e}thune} et~al.}{2016}]{bethune16}
{B{\'e}thune} W.,  {Lesur} G.,   {Ferreira} J.,  2016, \mn@doi [\aap]
  {10.1051/0004-6361/201527874}, \href
  {https://ui.adsabs.harvard.edu/abs/2016A&A...589A..87B} {589, A87}

\bibitem[\protect\citeauthoryear{{B{\'e}thune}, {Lesur}  \&
  {Ferreira}}{{B{\'e}thune} et~al.}{2017}]{bethune17}
{B{\'e}thune} W.,  {Lesur} G.,   {Ferreira} J.,  2017, \mn@doi [\aap]
  {10.1051/0004-6361/201630056}, \href
  {https://ui.adsabs.harvard.edu/abs/2017A&A...600A..75B} {600, A75}

\bibitem[\protect\citeauthoryear{{Blaes} \& {Balbus}}{{Blaes} \&
  {Balbus}}{1994}]{bb94}
{Blaes} O.~M.,  {Balbus} S.~A.,  1994, \mn@doi [\apj] {10.1086/173634}, \href
  {https://ui.adsabs.harvard.edu/abs/1994ApJ...421..163B} {421, 163}
  
\bibitem[\protect\citeauthoryear{{Cui} \& {Bai}}{{Cui} \& {Bai}}{2020}]{cb20}
{Cui} C.,  {Bai} X.-N.,  2020, \mn@doi [\apj] {10.3847/1538-4357/ab7194}, \href
  {https://ui.adsabs.harvard.edu/abs/2020ApJ...891...30C} {891, 30}

\bibitem[\protect\citeauthoryear{{Cui} \& {Lin}}{{Cui} \& {Lin}}{2021}]{cl21}
{Cui} C.,  {Lin} M.-K.,  2021, \mn@doi [\mnras] {10.1093/mnras/stab1511}, \href
  {https://ui.adsabs.harvard.edu/abs/2021MNRAS.505.2983C} {505, 2983}

\bibitem[\protect\citeauthoryear{d'Alessio et al.}{1998}]{dalessio}
{d'Alessio} P.,  {Cant\"o} J., {Calvet} N., {Lizano} S.  1998, \mn@doi [\apj] {10.1086/305702}, \href
  {https://ui.adsabs.harvard.edu/abs/1998ApJ...500..411D} {500, 411}


\bibitem[\protect\citeauthoryear{{Desch}}{{Desch}}{2004}]{desch04}
{Desch} S.~J.,  2004, \mn@doi [\apj] {10.1086/392527}, \href
  {https://ui.adsabs.harvard.edu/abs/2004ApJ...608..509D} {608, 509}

\bibitem[\protect\citeauthoryear{{Flaherty}, {Hughes}, {Rosenfeld}, {Andrews},
  {Chiang}, {Simon}, {Kerzner}  \& {Wilner}}{{Flaherty}
  et~al.}{2015}]{flaherty15}
{Flaherty} K.~M.,  {Hughes} A.~M.,  {Rosenfeld} K.~A.,  {Andrews} S.~M.,
  {Chiang} E.,  {Simon} J.~B.,  {Kerzner} S.,   {Wilner} D.~J.,  2015, \mn@doi
  [\apj] {10.1088/0004-637X/813/2/99}, \href
  {https://ui.adsabs.harvard.edu/abs/2015ApJ...813...99F} {813, 99}

\bibitem[\protect\citeauthoryear{{Flaherty} et~al.,}{{Flaherty}
  et~al.}{2017}]{flaherty17}
{Flaherty} K.~M.,  et~al., 2017, \mn@doi [\apj] {10.3847/1538-4357/aa79f9},
  \href {https://ui.adsabs.harvard.edu/abs/2017ApJ...843..150F} {843, 150}

\bibitem[\protect\citeauthoryear{{Flaherty}, {Hughes}, {Teague}, {Simon},
  {Andrews}  \& {Wilner}}{{Flaherty} et~al.}{2018}]{flaherty18}
{Flaherty} K.~M.,  {Hughes} A.~M.,  {Teague} R.,  {Simon} J.~B.,  {Andrews}
  S.~M.,   {Wilner} D.~J.,  2018, \mn@doi [\apj] {10.3847/1538-4357/aab615},
  \href {https://ui.adsabs.harvard.edu/abs/2018ApJ...856..117F} {856, 117}

\bibitem[\protect\citeauthoryear{{Flock}, {Turner}, {Nelson}, {Lyra}, {Manger}
  \& {Klahr}}{{Flock} et~al.}{2020}]{flock20}
{Flock} M.,  {Turner} N.~J.,  {Nelson} R.~P.,  {Lyra} W.,  {Manger} N.,
  {Klahr} H.,  2020, \mn@doi [\apj] {10.3847/1538-4357/ab9641}, \href
  {https://ui.adsabs.harvard.edu/abs/2020ApJ...897..155F} {897, 155}

\bibitem[\protect\citeauthoryear{{Fricke}}{{Fricke}}{1968}]{fricke68}
{Fricke} K.,  1968, \zap, \href
  {https://ui.adsabs.harvard.edu/abs/1968ZA.....68..317F} {68, 317}

\bibitem[\protect\citeauthoryear{{Fromang} \& {Lesur}}{{Fromang} \&
  {Lesur}}{2019}]{fl19}
{Fromang} S.,  {Lesur} G.,  2019, in EAS Publications Series. p.~391,
  \mn@doi{10.1051/eas/1982035}

\bibitem[\protect\citeauthoryear{{Fukuhara}, {Okuzumi}  \& {Ono}}{{Fukuhara}
  et~al.}{2021}]{fukuhara14}
{Fukuhara} Y.,  {Okuzumi} S.,   {Ono} T.,  2021, \mn@doi [\apj]
  {10.3847/1538-4357/abfe5c}, \href
  {https://ui.adsabs.harvard.edu/abs/2021ApJ...914..132F} {914, 132}

\bibitem[\protect\citeauthoryear{{Goldreich} \& {Schubert}}{{Goldreich} \&
  {Schubert}}{1967}]{gs67}
{Goldreich} P.,  {Schubert} G.,  1967, \mn@doi [\apj] {10.1086/149360}, \href
  {https://ui.adsabs.harvard.edu/abs/1967ApJ...150..571G} {150, 571}

\bibitem[\protect\citeauthoryear{{Gressel}, {Turner}, {Nelson}  \&
  {McNally}}{{Gressel} et~al.}{2015}]{gressel15}
{Gressel} O.,  {Turner} N.~J.,  {Nelson} R.~P.,   {McNally} C.~P.,  2015,
  \mn@doi [\apj] {10.1088/0004-637X/801/2/84}, \href
  {https://ui.adsabs.harvard.edu/abs/2015ApJ...801...84G} {801, 84}

\bibitem[\protect\citeauthoryear{{Guilloteau}, {Dutrey}, {Pi{\'e}tu}  \&
  {Boehler}}{{Guilloteau} et~al.}{2011}]{guill2011}
{Guilloteau} S.,  {Dutrey} A.,  {Pi{\'e}tu} V.,   {Boehler} Y.,  2011, \mn@doi
  [\aap] {10.1051/0004-6361/201015209}, \href
  {https://ui.adsabs.harvard.edu/abs/2011A&A...529A.105G} {529, A105}

\bibitem[\protect\citeauthoryear{{Hayashi}}{{Hayashi}}{1981}]{hayashi81}
{Hayashi} C.,  1981, \mn@doi [Progress of Theoretical Physics Supplement]
  {10.1143/PTPS.70.35}, \href
  {https://ui.adsabs.harvard.edu/abs/1981PThPS..70...35H} {70, 35}

\bibitem[\protect\citeauthoryear{{Isella}, {Carpenter}  \& {Sargent}}{{Isella}
  et~al.}{2009}]{isella2009}
{Isella} A.,  {Carpenter} J.~M.,   {Sargent} A.~I.,  2009, \mn@doi [\apj]
  {10.1088/0004-637X/701/1/260}, \href
  {https://ui.adsabs.harvard.edu/abs/2009ApJ...701..260I} {701, 260}

\bibitem[\protect\citeauthoryear{{Knobloch} \& {Spruit}}{{Knobloch} \&
  {Spruit}}{1982}]{ks82}
{Knobloch} E.,  {Spruit} H.~C.,  1982, \aap, \href
  {https://ui.adsabs.harvard.edu/abs/1982A&A...113..261K} {113, 261}
  
  \bibitem[\protect\citeauthoryear{{Knobloch} \& {Spruit}}{{Knobloch} \&
  {Spruit}}{1986}]{ks86}
{Knobloch} E.,  {Spruit} H.~C.,  1986, \aap, \href
  {https://ui.adsabs.harvard.edu/abs/1986A&A...166..359K} {166, 359}

\bibitem[\protect\citeauthoryear{{Krapp}, {Gressel}, {Ben{\'\i}tez-Llambay},
  {Downes}, {Mohandas}  \& {Pessah}}{{Krapp} et~al.}{2018}]{krapp18}
{Krapp} L.,  {Gressel} O.,  {Ben{\'\i}tez-Llambay} P.,  {Downes} T.~P.,
  {Mohandas} G.,   {Pessah} M.~E.,  2018, \mn@doi [\apj]
  {10.3847/1538-4357/aadcf0}, \href
  {https://ui.adsabs.harvard.edu/abs/2018ApJ...865..105K} {865, 105}

\bibitem[\protect\citeauthoryear{{Kratter} \& {Lodato}}{{Kratter} \&
  {Lodato}}{2016}]{kratter16}
{Kratter} K.,  {Lodato} G.,  2016, \mn@doi [\araa] {10.1146/annurev-astro-081915-023307},
  \href {https://ui.adsabs.harvard.edu/abs/2016ARA&A..54..271K} {54, 271}

\bibitem[\protect\citeauthoryear{{Kunz}}{{Kunz}}{2008}]{kunz08}
{Kunz} M.~W.,  2008, \mn@doi [\mnras] {10.1111/j.1365-2966.2008.12928.x}, \href
  {http://adsabs.harvard.edu/abs/2008MNRAS.385.1494K} {385, 1494}

\bibitem[\protect\citeauthoryear{{Kunz} \& {Balbus}}{{Kunz} \&
  {Balbus}}{2004}]{kb04}
{Kunz} M.~W.,  {Balbus} S.~A.,  2004, \mn@doi [\mnras]
  {10.1111/j.1365-2966.2004.07383.x}, \href
  {https://ui.adsabs.harvard.edu/abs/2004MNRAS.348..355K} {348, 355}

\bibitem[\protect\citeauthoryear{{Kunz} \& {Lesur}}{{Kunz} \&
  {Lesur}}{2013}]{kl13}
{Kunz} M.~W.,  {Lesur} G.,  2013, \mn@doi [\mnras] {10.1093/mnras/stt1171},
  \href {http://adsabs.harvard.edu/abs/2013MNRAS.434.2295K} {434, 2295}

\bibitem[\protect\citeauthoryear{{Kunz} \& {Mouschovias}}{{Kunz} \&
  {Mouschovias}}{2009}]{km09}
{Kunz} M.~W.,  {Mouschovias} T.~C.,  2009, \mn@doi [\apj]
  {10.1088/0004-637X/693/2/1895}, \href
  {https://ui.adsabs.harvard.edu/abs/2009ApJ...693.1895K} {693, 1895}

\bibitem[\protect\citeauthoryear{{Latter} \& {Papaloizou}}{{Latter} \&
  {Papaloizou}}{2017}]{lp17}
{Latter} H.~N.,  {Papaloizou} J.,  2017, \mn@doi [\mnras]
  {10.1093/mnras/stx2038}, \href
  {https://ui.adsabs.harvard.edu/abs/2017MNRAS.472.1432L} {472, 1432}

\bibitem[\protect\citeauthoryear{{Latter} \& {Papaloizou}}{{Latter} \&
  {Papaloizou}}{2018}]{lp18}
{Latter} H.~N.,  {Papaloizou} J.,  2018, \mn@doi [\mnras]
  {10.1093/mnras/stx3031}, \href
  {https://ui.adsabs.harvard.edu/abs/2018MNRAS.474.3110L} {474, 3110} (LP18)

\bibitem[\protect\citeauthoryear{{Latter}, {Fromang}  \& {Gressel}}{{Latter}
  et~al.}{2010}]{Latter2010}
{Latter} H.~N.,  {Fromang} S.,   {Gressel} O.,  2010, \mn@doi [\mnras]
  {10.1111/j.1365-2966.2010.16759.x}, \href
  {https://ui.adsabs.harvard.edu/abs/2010MNRAS.406..848L} {406, 848}

\bibitem[\protect\citeauthoryear{{Lesur}}{{Lesur}}{2020}]{lesur20}
{Lesur} G.,  2020, arXiv e-prints, \href
  {https://ui.adsabs.harvard.edu/abs/2020arXiv200715967L} {arXiv:2007.15967}

\bibitem[\protect\citeauthoryear{{Lesur}, {Hennebelle}  \& {Fromang}}{{Lesur}
  et~al.}{2015}]{lesur15}
{Lesur} G.,  {Hennebelle} P.,   {Fromang} S.,  2015, \mn@doi [\aap]
  {10.1051/0004-6361/201526734}, \href
  {https://ui.adsabs.harvard.edu/abs/2015A&A...582L...9L} {582, L9}

\bibitem[\protect\citeauthoryear{{Lesur}, {Kunz}  \& {Fromang}}{{Lesur}
  et~al.}{2014}]{lkf14}
{Lesur} G.,  {Kunz} M.~W.,   {Fromang} S.,  2014, \mn@doi [\aap]
  {10.1051/0004-6361/201423660}, \href
  {https://ui.adsabs.harvard.edu/abs/2014A&A...566A..56L} {566, A56}

\bibitem[\protect\citeauthoryear{{Leung} \& {Ogilvie}}{{Leung} \&
  {Ogilvie}}{2019}]{lo19}
{Leung} P. K.~C.,  {Ogilvie} G.~I.,  2019, \mn@doi [\mnras]
  {10.1093/mnras/stz1620}, \href
  {https://ui.adsabs.harvard.edu/abs/2019MNRAS.487.5155L} {487, 5155}

\bibitem[\protect\citeauthoryear{{Lin} \& {Youdin}}{{Lin} \&
  {Youdin}}{2015}]{ly15}
{Lin} M.-K.,  {Youdin} A.~N.,  2015, \mn@doi [\apj]
  {10.1088/0004-637X/811/1/17}, \href
  {https://ui.adsabs.harvard.edu/abs/2015ApJ...811...17L} {811, 17}

\bibitem[\protect\citeauthoryear{{Lyra} \& {Umurhan}}{{Lyra} \&
  {Umurhan}}{2019}]{lu19}
{Lyra} W.,  {Umurhan} O.~M.,  2019, \mn@doi [\pasp] {10.1088/1538-3873/aaf5ff},
  \href {https://ui.adsabs.harvard.edu/abs/2019PASP..131g2001L} {131, 072001}

\bibitem[\protect\citeauthoryear{{Malygin}, {Klahr}, {Semenov}, {Henning}  \&
  {Dullemond}}{{Malygin} et~al.}{2017}]{malygin17}
{Malygin} M.~G.,  {Klahr} H.,  {Semenov} D.,  {Henning} T.,   {Dullemond}
  C.~P.,  2017, \mn@doi [\aap] {10.1051/0004-6361/201629933}, \href
  {https://ui.adsabs.harvard.edu/abs/2017A&A...605A..30M} {605, A30}

\bibitem[\protect\citeauthoryear{{Masson}, {Chabrier}, {Hennebelle}, {Vaytet}
  \& {Commer{\c{c}}on}}{{Masson} et~al.}{2016}]{masson16}
{Masson} J.,  {Chabrier} G.,  {Hennebelle} P.,  {Vaytet} N.,
  {Commer{\c{c}}on} B.,  2016, \mn@doi [\aap] {10.1051/0004-6361/201526371},
  \href {https://ui.adsabs.harvard.edu/abs/2016A&A...587A..32M} {587, A32}

\bibitem[\protect\citeauthoryear{{Nelson}, {Gressel}  \& {Umurhan}}{{Nelson}
  et~al.}{2013}]{nelson13}
{Nelson} R.~P.,  {Gressel} O.,   {Umurhan} O.~M.,  2013, \mn@doi [\mnras]
  {10.1093/mnras/stt1475}, \href
  {https://ui.adsabs.harvard.edu/abs/2013MNRAS.435.2610N} {435, 2610}

\bibitem[\protect\citeauthoryear{{Ogilvie}}{{Ogilvie}}{1997}]{ogilvie97}
{Ogilvie} G.~I.,  1997, \mn@doi [\mnras] {10.1093/mnras/288.1.63}, \href
  {https://ui.adsabs.harvard.edu/abs/1997MNRAS.288...63O} {288, 63}

\bibitem[\protect\citeauthoryear{{Ogilvie} \& {Livio}}{{Ogilvie} \&
  {Livio}}{2001}]{ol01}
{Ogilvie} G.~I.,  {Livio} M.,  2001, \mn@doi [\apj] {10.1086/320637}, \href
  {https://ui.adsabs.harvard.edu/abs/2001ApJ...553..158O} {553, 158}

\bibitem[\protect\citeauthoryear{{Pandey} \& {Wardle}}{{Pandey} \&
  {Wardle}}{2008}]{pw08}
{Pandey} B.~P.,  {Wardle} M.,  2008, \mn@doi [\mnras]
  {10.1111/j.1365-2966.2008.12998.x}, \href
  {https://ui.adsabs.harvard.edu/abs/2008MNRAS.385.2269P} {385, 2269}
  
\bibitem[\protect\citeauthoryear{{Pandey} \& {Wardle}}{{Pandey} \&
  {Wardle}}{2012}]{pw12}
{Pandey} B.~P.,  {Wardle} M.,  2012, \mn@doi [\mnras]
  {10.1111/j.1365-2966.2012.20799.x}, \href
  {https://ui.adsabs.harvard.edu/abs/2012MNRAS.423..222P} {423, 222}

\bibitem[\protect\citeauthoryear{{Pfeil} \& {Klahr}}{{Pfeil} \&
  {Klahr}}{2019}]{pk19}
{Pfeil} T.,  {Klahr} H.,  2019, \mn@doi [\apj] {10.3847/1538-4357/aaf962},
  \href {https://ui.adsabs.harvard.edu/abs/2019ApJ...871..150P} {871, 150}

\bibitem[\protect\citeauthoryear{{Pfeil} \& {Klahr}}{{Pfeil} \&
  {Klahr}}{2021}]{pk21}
{Pfeil} T.,  {Klahr} H.,  2021, \mn@doi [\apj] {10.3847/1538-4357/ac0054},
  \href {https://ui.adsabs.harvard.edu/abs/2021ApJ...915..130P} {915, 130}

\bibitem[\protect\citeauthoryear{{Richard}, {Nelson}  \& {Umurhan}}{{Richard}
  et~al.}{2016}]{richard16}
{Richard} S.,  {Nelson} R.~P.,   {Umurhan} O.~M.,  2016, \mn@doi [\mnras]
  {10.1093/mnras/stv2898}, \href
  {https://ui.adsabs.harvard.edu/abs/2016MNRAS.456.3571R} {456, 3571}

\bibitem[\protect\citeauthoryear{{Salmeron}, {K{\"o}nigl}  \&
  {Wardle}}{{Salmeron} et~al.}{2011}]{salmeron11}
{Salmeron} R.,  {K{\"o}nigl} A.,   {Wardle} M.,  2011, \mn@doi [\mnras]
  {10.1111/j.1365-2966.2010.17974.x}, \href
  {https://ui.adsabs.harvard.edu/abs/2011MNRAS.412.1162S} {412, 1162}

\bibitem[\protect\citeauthoryear{{Simon}, {Lesur}, {Kunz}  \&
  {Armitage}}{{Simon} et~al.}{2015}]{simon15}
{Simon} J.~B.,  {Lesur} G.,  {Kunz} M.~W.,   {Armitage} P.~J.,  2015, \mn@doi
  [\mnras] {10.1093/mnras/stv2070}, \href
  {https://ui.adsabs.harvard.edu/abs/2015MNRAS.454.1117S} {454, 1117}

\bibitem[\protect\citeauthoryear{{Stoll} \& {Kley}}{{Stoll} \&
  {Kley}}{2014}]{sk14}
{Stoll} M. H.~R.,  {Kley} W.,  2014, \mn@doi [\aap]
  {10.1051/0004-6361/201424114}, \href
  {https://ui.adsabs.harvard.edu/abs/2014A&A...572A..77S} {572, A77}

\bibitem[\protect\citeauthoryear{{Stoll} \& {Kley}}{{Stoll} \&
  {Kley}}{2016}]{sk16}
{Stoll} M. H.~R.,  {Kley} W.,  2016, \mn@doi [\aap]
  {10.1051/0004-6361/201527716}, \href
  {https://ui.adsabs.harvard.edu/abs/2016A&A...594A..57S} {594, A57}

\bibitem[\protect\citeauthoryear{{Stoll}, {Kley}  \& {Picogna}}{{Stoll}
  et~al.}{2017}]{stoll17}
{Stoll} M. H.~R.,  {Kley} W.,   {Picogna} G.,  2017, \mn@doi [\aap]
  {10.1051/0004-6361/201630226}, \href
  {https://ui.adsabs.harvard.edu/abs/2017A&A...599L...6S} {599, L6}

\bibitem[\protect\citeauthoryear{{Suriano}, {Li}, {Krasnopolsky}  \&
  {Shang}}{{Suriano} et~al.}{2018}]{suriano18}
{Suriano} S.~S.,  {Li} Z.-Y.,  {Krasnopolsky} R.,   {Shang} H.,  2018, \mn@doi
  [\mnras] {10.1093/mnras/sty717}, \href
  {https://ui.adsabs.harvard.edu/abs/2018MNRAS.477.1239S} {477, 1239}

\bibitem[\protect\citeauthoryear{{Suriano}, {Li}, {Krasnopolsky}, {Suzuki}  \&
  {Shang}}{{Suriano} et~al.}{2019}]{suriano19}
{Suriano} S.~S.,  {Li} Z.-Y.,  {Krasnopolsky} R.,  {Suzuki} T.~K.,   {Shang}
  H.,  2019, \mn@doi [\mnras] {10.1093/mnras/sty3502}, \href
  {https://ui.adsabs.harvard.edu/abs/2019MNRAS.484..107S} {484, 107}

\bibitem[\protect\citeauthoryear{{Tazzari} et~al.,}{{Tazzari}
  et~al.}{2017}]{tazzari2017}
{Tazzari} M.,  et~al., 2017, \mn@doi [\aap] {10.1051/0004-6361/201730890},
  \href {https://ui.adsabs.harvard.edu/abs/2017A&A...606A..88T} {606, A88}

\bibitem[\protect\citeauthoryear{{Turner}, {Fromang}, {Gammie}, {Klahr},
  {Lesur}, {Wardle}  \& {Bai}}{{Turner} et~al.}{2014}]{turner14}
{Turner} N.~J.,  {Fromang} S.,  {Gammie} C.,  {Klahr} H.,  {Lesur} G.,
  {Wardle} M.,   {Bai} X.~N.,  2014, in {Beuther} H.,  {Klessen} R.~S.,
  {Dullemond} C.~P.,   {Henning} T.,  eds, Protostars and Planets VI. p.~411, 
  \mn@doi{10.2458/azu\_uapress\_9780816531240-ch018}

\bibitem[\protect\citeauthoryear{{Umebayashi} \& {Nakano}}{{Umebayashi} \&
  {Nakano}}{1980}]{un80}
{Umebayashi} T.,  {Nakano} T.,  1980, \pasj, \href
  {https://ui.adsabs.harvard.edu/abs/1980PASJ...32..405U} {32, 405}

\bibitem[\protect\citeauthoryear{{Umebayashi} \& {Nakano}}{{Umebayashi} \&
  {Nakano}}{2009}]{un09}
{Umebayashi} T.,  {Nakano} T.,  2009, \mn@doi [\apj]
  {10.1088/0004-637X/690/1/69}, \href
  {https://ui.adsabs.harvard.edu/abs/2009ApJ...690...69U} {690, 69}

\bibitem[\protect\citeauthoryear{{Urpin} \& {Brandenburg}}{{Urpin} \&
  {Brandenburg}}{1998}]{ub98}
{Urpin} V.,  {Brandenburg} A.,  1998, \mn@doi [\mnras]
  {10.1046/j.1365-8711.1998.01118.x}, \href
  {https://ui.adsabs.harvard.edu/abs/1998MNRAS.294..399U} {294, 399}

\bibitem[\protect\citeauthoryear{{Vlemmings} et~al.,}{{Vlemmings}
  et~al.}{2019}]{vlemmings19}
{Vlemmings} W.~H.~T.,  et~al., 2019, \mn@doi [\aap]
  {10.1051/0004-6361/201935459}, \href
  {https://ui.adsabs.harvard.edu/abs/2019A&A...624L...7V} {624, L7}

\bibitem[\protect\citeauthoryear{{Wardle}}{{Wardle}}{1999}]{wardle99}
{Wardle} M.,  1999, \mn@doi [\mnras] {10.1046/j.1365-8711.1999.02670.x}, \href
  {https://ui.adsabs.harvard.edu/abs/1999MNRAS.307..849W} {307, 849}

\bibitem[\protect\citeauthoryear{{Wardle} \& {Salmeron}}{{Wardle} \& {Salmeron}}{2012}]{ws12}
{Wardle} M., {Salmeron} R., 2012, \mn@doi [\mnras] {10.1111/j.1365-2966.2011.20022.x}, \href
  {https://ui.adsabs.harvard.edu/abs/2012MNRAS.422.2737W} {422, 2737}

\bibitem[\protect\citeauthoryear{{Xu} \& {Kunz}}{{Xu} \& {Kunz}}{2021}]{xk21b}
{Xu} W.,  {Kunz} M.~W.,  2021, \mn@doi [\mnras] {10.1093/mnras/stab2715}, \href
  {https://ui.adsabs.harvard.edu/abs/2021MNRAS.508.2142X} {508, 2142}

\makeatother
\end{thebibliography}
\end{document}